\documentclass[a4paper,fleqn]{cas-dc}

\usepackage[numbers]{natbib}
\usepackage[ruled,vlined,linesnumbered]{algorithm2e}

\usepackage{graphicx}
\usepackage{placeins}
\usepackage{tikz}
\def\tsc#1{\csdef{#1}{\textsc{\lowercase{#1}}\xspace}}
\tsc{WGM}
\tsc{QE}
\tsc{EP}
\tsc{PMS}
\tsc{BEC}
\tsc{DE}

\begin{document}
\let\WriteBookmarks\relax
\def\floatpagepagefraction{1}
\def\textpagefraction{.001}
\shorttitle{Eliminating Premature Termination in Multihop Rendezvous for Cognitive Radio based Emergency Response Network}
\shortauthors{Z. Ali et~al.}

\title [mode = title]{Eliminating Premature Termination in Multihop Rendezvous for Cognitive Radio based Emergency Response Network
}                      

\tnotetext[1]{This Research has been supported by the Atlantic Technological University, Ireland under the Postgraduate Research Training Program in Modeling and Computation for Health and Society (MOCHAS PRTP).}

\author[1]{Zahid Ali}[type=editor,
                        orcid=0000-0002-6066-6133
                        ]
\ead{za4zahidali@gmail.com}


\affiliation[1]{
	organization={Department of Computing, Atlantic Technological University,},
                       state={Donegal},
                country={Ireland}}

\author[2]{Saritha Unnikrishnan}[type=editor,
]


\affiliation[2]{
	organization={Department of Computing \& Electronic Engineering, Atlantic Technological University,},
	state={Sligo},
	country={Ireland}}

\author[1]{Eoghan Furey}[type=editor,
]

\author[3]{Ian McLoughlin}[type=editor,
]

\author[1]{Saim Ghafoor}[type=editor,
]

\affiliation[3]{
	organization={Department of Computer Science \& Applied Physics, Atlantic Technological University,},
	state={Galway},
	country={Ireland}}

\cortext[cor1]{Corresponding author}

\begin{abstract}
In post-disaster environments, damaged communication infrastructure severely limits coordination among emergency response teams. Cognitive radio networks (CRNs) enable rapidly deployable communication by allowing nodes to opportunistically access available spectrum. However, existing multihop rendezvous protocols typically rely on N-1 termination conditions, which can lead to premature termination, resulting in incomplete neighbour discovery and invalid network topology formation. This work identifies this limitation as a previously overlooked issue in multihop rendezvous protocols. This paper proposes a Multihop Reliable Dual-Modular Clock Algorithm (MR-DMCA) that eliminates premature termination and ensures reliable network formation. The proposed protocol introduces a coordinate-assisted neighbour validation mechanism and an autonomous termination strategy that guarantees complete neighbour and topology discovery before protocol termination. Although implemented within MR-DMCA, the proposed validation and termination approach is applicable to a wider class of multihop rendezvous protocols. Extensive simulations demonstrate that, in a worst-case scalable scenario with 20 nodes and 20 channels under high primary radio activity (m=2), MR-DMCA achieves 100\% accurate neighbour and topology discovery while reducing rendezvous time by up to 76\%, 37\%, and 17\% compared with baseline protocols. The results highlight that addressing premature termination is critical for reliable multihop rendezvous in cognitive radio-based emergency communication networks.

\end{abstract}




\maketitle

\section{Introduction}
Cognitive Radio Networks (CRNs) enable dynamic spectrum access by allowing secondary users to opportunistically utilise underused spectrum bands \cite{4}. This capability makes CRNs particularly suitable for rapidly deployable communication systems where fixed infrastructure may be unavailable, such as temporary or emergency communication environments \cite{1,2,3}. In such distributed networks, nodes must autonomously discover neighbouring nodes and establish connectivity over dynamically available channels. Rendezvous protocols play a fundamental role in CRNs by enabling two nodes to meet on a common channel at the same time to exchange control information and initiate communication. While extensive research has been conducted on rendezvous algorithms, most existing approaches \cite{5, Ali2025, 6, Ali2026} primarily focus on minimising rendezvous delay and improving channel utilisation efficiency. Furthermore, many studies \cite{5, Ali2025} assume single-hop environments, whereas practical CRNs often operate in multihop scenarios, where nodes must discover neighbours and form a complete network topology \cite{6, Ali2026}.

In multihop rendezvous protocols, neighbour discovery and topology formation typically proceed through repeated rendezvous attempts across multiple channels. However, the termination mechanisms used in existing approaches may stop the discovery process prematurely when predefined stopping conditions are met. In particular, some approaches \cite{6, Ali2026} rely on the assumption that once a node discovers N-1 neighbours, where N represents the total number of nodes in the network, the discovery process can be safely terminated. In multihop environments, however, nodes often lack global knowledge of network size and connectivity, and the N-1 discovery condition may be incorrectly satisfied, resulting in incomplete neighbour discovery and inaccurate topology construction. To the best of our knowledge, this limitation has not been explicitly addressed in existing multihop rendezvous protocols, where termination is typically assumed to imply complete and correct topology discovery. Although implemented within MR-DMCA, the proposed validation and termination mechanism is generic and can be applied to other multihop rendezvous protocols.

To address this problem, this paper proposes a Multihop Reliable Dual Modular Clock Algorithm (MR-DMCA) that incorporates a coordinate-assisted identification phase for validating discovered neighbours. The proposed mechanism enables nodes to verify topology completeness before terminating the discovery process, thereby preventing premature termination and improving the reliability of multihop rendezvous. The proposed MR-DMCA builds upon our previous multihop rendezvous protocol (M-DMCA) \cite{Ali2026}, which does not include the IDN phase or topology validation mechanism. MR-DMCA introduces these enhancements to prevent premature termination and ensure accurate topology discovery.

The main contributions of this work are summarised as follows:
\begin{enumerate}
	\item This work identifies and analyses a previously overlooked premature termination problem in existing multihop rendezvous protocols that rely on the N-1 termination condition. Such termination assumptions can result in incomplete neighbour discovery and inaccurate topology formation in cognitive radio networks.
	\item A coordinate-assisted neighbour validation mechanism is introduced to verify discovered neighbours and ensure accurate topology formation before protocol termination.
	\item A Multihop Reliable Dual-Modular Clock Algorithm (MR-DMCA) is proposed, which integrates the coordinate assisted validation with a controlled termination strategy to ensure correct and complete topology discovery. 
	\item We introduce Average Topology Match (ATM) and Post-Termination Discovery Delay (PTDD) metrics to quantify topology correctness and the cost of achieving reliable discovery.
\end{enumerate}	

Unlike prior work that focuses primarily on rendezvous efficiency, this study emphasises topology correctness and reliable termination as essential requirements for multihop cognitive radio networks.

The rest of a paper is organised as, the Related work is explained in Section II, Preliminaries and system model discussed network and PR activity, dual clock,  neighbour discovery and termination condition in Section III. Multihop Reliable DMCA is presented in Section IV. Section V, explains Topology validation,  then performance evaluations in Section VI. Finally, the conclusion is given in Section VII.

\section{Related Work}
Multihop rendezvous in cognitive radio networks (CRNs) has been extensively studied to enable distributed neighbour discovery and network formation under dynamic spectrum conditions \cite{gg,hh}. Early approaches \cite{li2017} include sequence-based protocols such as Jump-Stay \cite{Hliu2012,paul2016, yu2015}, Galois-field based \cite{g1,g2}, Quorum-based \cite{chuang2015, chao2016,sheu2016,zhang2014}, Matrix-based \cite{huang2017,chang2015}, number-theoretic schemes \cite{sahoo2016, chao2016efficient, chang2014} and combinatorics\cite{yang2016}, which generate predetermined channel hopping sequences to guarantee rendezvous. While these methods provide bounded time-to-rendezvous (TTR), their reliance on fixed sequences limits adaptability in highly dynamic or unpredictable environments, such as post-disaster scenarios.

Clock-based protocols were introduced to address this limitation by selecting channels based on sensed availability rather than predefined sequences. Notable examples include Random Channel Selection (RCS), Modular Clock Algorithm (MCA), Modified MCA (MMCA)\cite{5}, EMCA\cite{6}, DMCA\cite{Ali2025} and M-DMCA\cite{Ali2026}. These protocols improve flexibility and support multihop coordination without relying on predetermined sequences, making them better suited for emergency or disaster-response networks. However, they primarily focus on reducing TTR and rely on termination rules based on N-1 neighbour discovery condition, assuming that network topology is complete once this condition is satisfied \cite{6, Ali2026}. In practice, even when the N-1 condition is met, the resulting topology may be inaccurate or incomplete, which can negatively affect subsequent communication phases such as routing and data dissemination. 

While EMCA and M-DMCA perform handshake-based neighbour exchanges, they do not include additional verification mechanism, such as coordinate-assisted validation, to ensure topology correctness \cite{6, Ali2026}. Most existing studies implicitly assume that successful rendezvous completion implies correct topology formation, without explicitly validating the accuracy of discovered neighbour relationships. However, existing studies do not explicitly address the issue of topology correctness under the N-1 termination condition, particularly when neighbour information is obtained indirectly through other nodes rather than a direct handshake. This motivates the proposed MR-DMCA protocol, which ensures accurate neighbour discovery, reliable topology formation, and controlled termination under dynamic spectrum conditions.    

Unlike existing multihop rendezvous protocols, MR-DMCA explicitly addresses premature termination in terms of topology correctness and introduces a coordinate-assisted neighbour validation phase, ensuring complete and reliable topology formation. To our knowledge, this is the first work to combine termination correctness analysis with coordinate-based validation in multihop cognitive radio rendezvous. 

\section{Preliminaries and System Model}
This section presents the system assumptions and operational models used in the proposed protocol. The network model, primary radio (PR) activity model, dual channel selection mechanism, and timeslot structure follow the framework introduced in our previous work \cite{Ali2026}. For completeness, a brief overview is provided. The key extensions introduced in this work, including coordinate-assisted neighbour validation and controlled termination, are described in the subsequent sections.

\subsection{Network Model}
We consider a multihop cognitive radio network composed of randomly deployed static secondary users (SUs), modelled as a connected graph to enable multihop communication and complete topology discovery. Each node is equipped with a single radio interface capable of performing spectrum sensing and opportunistic channel access. Due to spatial variations and primary radio activity, the set of available channels may vary across nodes. The network operates in a slotted system with synchronised timeslots. Each timeslot has a duration  T, which is sufficient to support spectrum sensing and packet exchange between rendezvousing nodes. Following the framework adopted in \cite{rr, ss}, a timeslot can be divided into transmission and reception intervals, typically divided into two equal intervals of duration 
T/2, corresponding to transmission and reception phases. For channel sensing, an energy detection model is assumed for identifying available spectrum opportunities, although more advanced techniques such as cyclostationary feature detection may also be applied for primary user detection \cite{tt}. 

\subsection{Primary Radio Activity Model}
Primary radio activity is modelled using a continuous-time alternating ON/OFF Markov renewal process\cite{bk}, which captures the dynamic occupancy of licensed channels. In this model, each channel alternates between busy $T^i_\text{ON}$ and idle $T^i_\text{OFF}$ states with exponentially distributed durations. The channel utilisation $U_i$ represents the fraction of time a channel is occupied by a PR. 

\begin{equation}\label{eq:Ui}
	U^i =  \frac {E[T^i_{\text{ON}}]}{ E[T^i_{\text{ON}}] + E[T^i_{\text{OFF}}]} = \frac{\lambda_Y}{\lambda_X+\lambda_Y}
\end{equation}
The probabilities of a channel being in the ON or OFF state are determined by the corresponding transition rates, using the formulas.
\begin{equation}\label{eq:Pon}
	P_{ON}(t) =  \frac{\lambda_Y}{\lambda_X+\lambda_Y} - \frac{\lambda_Y}{\lambda_X+\lambda_Y}e^{-(\lambda_X+\lambda_Y)t}
\end{equation}
\begin{equation}\label{eq:Poff}
	P_{OFF}(t) =  \frac{\lambda_X}{\lambda_X+\lambda_Y} + \frac{\lambda_Y}{\lambda_X+\lambda_Y}e^{-(\lambda_X+\lambda_Y)t}
\end{equation}
The formulation of the PR activity model and its analytical expressions are provided in \cite{bk} and also in \cite{uu, bj, xx}, while in this work, the model is used to generate different PR activity levels by adjusting the channel rate parameters. With the above mentioned formulas, the zero and high PR activity levels are generated by channel rate parameters ($\lambda_X$ and $\lambda_Y$) of \cite{Ali2025} and \cite{Ali2026}.
\subsection{Dual Clock and Timeslot Structure}
\begin{figure}[htbp]
	\centering
	\includegraphics[width=0.7\linewidth]{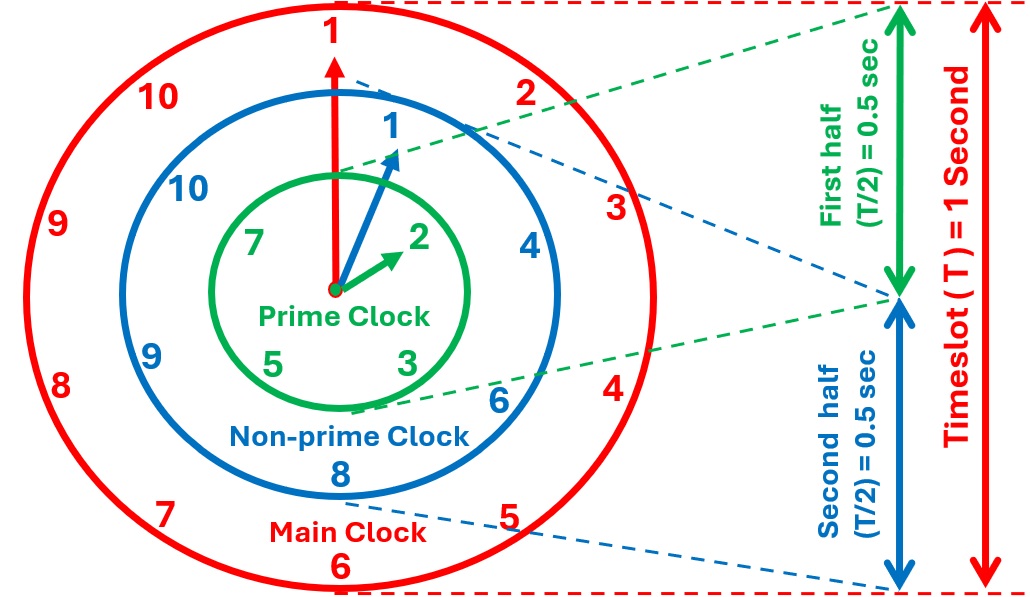} 
	\caption{Dual Clock Mechanism and Timeslot Division}
	\label{fig:IDN_clock}
\end{figure}

The protocol adopts the dual modular clock-based channel selection mechanism introduced in \cite{Ali2025, Ali2026}. Let m denote the number of available channels sensed by a node, represented as $CU_{i}$= $\{1, 2, 3, 4 ... m\}$. Each node maintains channel indices and performs channel hopping based on these indices. Unlike conventional clock-based approaches such as RCS, MCA, and EMCA that rely on a single clock for channel hopping, the dual modular clock mechanism divides the sensed channels into two independent sets based on their primality: prime channels and non-prime channels, as shown in Fig.~\ref{fig:IDN_clock}. Nodes perform channel hopping over these two channel sets using separate modular clocks within the same timeslot. Since the number of prime channels is typically smaller, the search space is reduced during the initial hopping phase, potentially improving rendezvous opportunities. If rendezvous is not achieved, nodes continue hopping over the non-prime channel set within the same slot. If one of the channel sets is empty, the protocol falls back to the main clock mechanism described in \cite{Ali2025, Ali2026}. 
\subsection{Neighbour Discovery}
In traditional multihop rendezvous protocols such as EMCA \cite{6} and M-DMCA \cite{Ali2026}, neighbour discovery is performed through a handshake-based information exchange. During this process, nodes exchange their neighbour information to construct a neighbour list. Typically, the neighbour list is composed of directly discovered neighbours, referred to as the Direct Neighbour List (DNL), and indirect neighbours, referred to as the Indirect Neighbour List (INL). Nodes in the DNL are discovered through successful direct handshakes, whereas nodes in the INL are inferred through information received from DNL nodes. Although this approach accelerates neighbour discovery, it may lead to inaccurate topology information because some neighbours are inferred indirectly without direct verification, which may result in incorrect classification of neighbours. Consequently, nodes may assume that the neighbour discovery process is complete even though some neighbours have not been directly validated. To improve the reliability of topology formation, the proposed approach incorporates a coordinate-assisted validation mechanism to verify neighbour relationships during the discovery process, as described in Section IV.

\subsection{Termination Condition}
Most existing multihop rendezvous protocols terminate the discovery phase once a node identifies $N-1$, neighbours, 
where N represents the total number of nodes in the network. This assumption implies a node has discovered all other nodes and that the topology is complete. However, when neighbour information is partially inferred through indirect exchanges rather than direct handshakes, the $N-1$ condition may be satisfied even though some neighbours have not been directly verified. As a result, nodes may terminate the rendezvous process with incomplete or inaccurate topology information, which can negatively affect subsequent communication stages such as routing, bandwidth allocation, and data dissemination. To address this limitation, the proposed MR-DMCA protocol introduces a coordinate-assisted validation mechanism and a controlled termination strategy to ensure that the discovered neighbours are correctly verified before allowing protocol termination.  
\begin{figure}[htbp]
	\centering
	\includegraphics[width=0.7\linewidth]{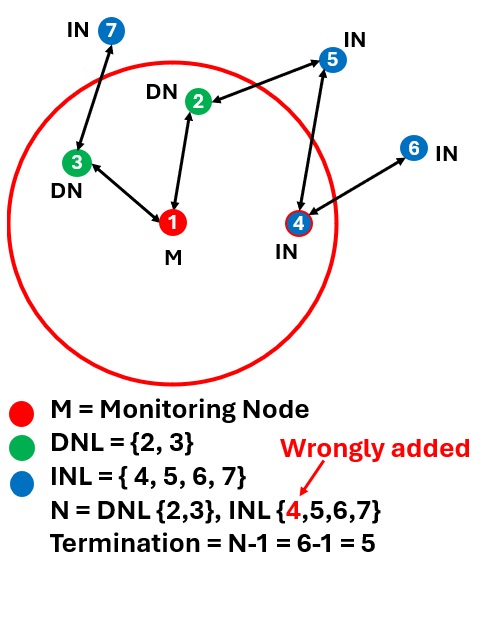} 
	\caption{Traditional termination }
	\label{fig:NlistaN}
\end{figure}

\section{Multihop Reliable Dual Modular Clock Algorithm (MR-DMCA)}
\begin{figure*}[ht] 
	\centering
	\includegraphics[width=0.852\textwidth]{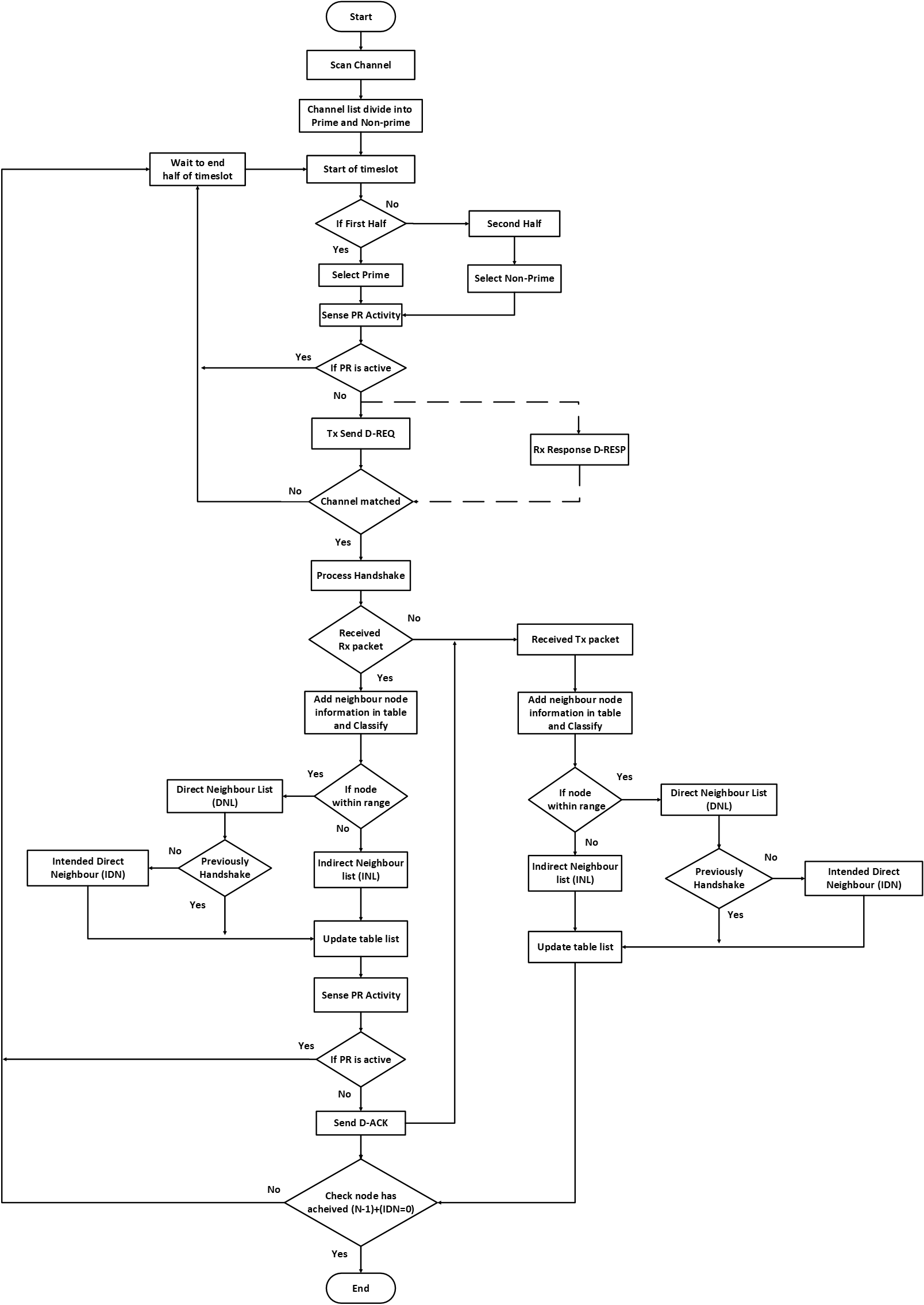}
	\caption{A complete flow chart of MR-DMCA}
	\label{fig:MR-DMCA2}
\end{figure*}

The Multihop Reliable Dual Modular Clock Algorithm (MR-DMCA) extends DMCA \cite{Ali2025} and M-DMCA \cite{Ali2026} to achieve reliable multihop rendezvous with validated topology discovery. MR-DMCA combines dual modular clock hopping with coordinate-assisted neighbour verification to address limitations of traditional N-1 termination and ensure complete and accurate neighbour discovery. Its pseudocode is shown in Algorithm 1. The operation of MR-DMCA is also illustrated in Fig.~\ref{fig:MR-DMCA2}, which shows the flow of dual clock hopping, coordinate-assisted neighbour validation, and controlled termination.

\subsection{Dual Clock Hopping}
Each node identifies its available channel set $m_i$  and splits it into prime channels $M_p$ and non-prime channels $N_p$. For each subset, the node determines the hopping sequence parameters to support dual rendezvous attempts within a single timeslot. In the first half of the timeslot, the node hops on $M_p$ channels list. If no prime channels are available, it hops onto the main clock $m_i$. Whereas, in the second half of the timeslot, the node hops onto the non-prime channels list. If none are available, it defaults to the main clock $m_i$. This dual clock operation, by hopping over prime and non-prime channel subsets, increases the likelihood of successful rendezvous in each timeslot while facilitating multihop neighbour discovery \cite{Ali2026}.
\subsection{Neighbour Classification}

Each node maintains a neighbour list divided into three categories:
\begin{enumerate}
	\item Direct Neighbour List (DNL): The list of nodes to which a successful handshake has been achieved and are verified to be within the communication range. 
	\item Indirect Neighbour List (INL): The list of nodes reported by other neighbours but not directly verified within the current node’s transmission range. 
	\item Intended Direct Neighbour (IDN): The list of nodes whose coordinates indicate they are within range, but for which a direct handshake has not yet occurred. 
\end{enumerate}
This classification enables nodes to track which neighbours are pending direct verification, preventing premature termination.
\subsection{Coordinate-Assisted Rendezvous and Handshake}
During each rendezvous attempt, nodes exchange neighbour information along with coordinate data using a three-way handshake:
\begin{enumerate}
	\item {Discovery Request (D-REQ)}: The initiating node broadcasts its ID, coordinates, and neighbour tables (DNL, INL, IDN).
	\item {Discovery Response (D-RESP)}: Receiving nodes update their neighbour tables. The nodes within transmission range that successfully complete the handshake are added to the DNL. The nodes which are outside the range will be added to INL. Nodes whose coordinates indicate they are within transmission range but for which the handshake is not yet completed are included in the IDN list. A D-RESP is sent back containing updated neighbour information.
	\item {Discovery Acknowledgment (D-ACK)}: The initiating node confirms receipt by sending a D-ACK message and updates its neighbour tables accordingly.  
\end{enumerate}
Before attempting rendezvous on any channel, each node first senses the PR activity. If the channel is occupied by a PR, the node defers transmission; otherwise, it proceeds to transmit the packet.

\subsection{Controlled Termination}
Rendezvous continues until each node satisfies two conditions:
\begin{itemize}
	\item The neighbour list is complete, i.e, all $N-1$ neighbours are identified $|DNL_i \cup INL_i| = N - 1$
	\item All intended direct neighbours have been successfully verified, i.e., the IDN set is empty $(IDN=\emptyset)$
\end{itemize}

If any IDN remains, the node continues rendezvous until a successful handshake occurs and the IDN node is promoted to DNL. Only after both conditions are met does the node terminate the rendezvous phase. This ensures complete and accurate topology formation, preventing premature termination and supporting reliable subsequent operations like routing, resource allocation, and data dissemination.

In MR-DMCA each node first identifies its available channel set and divides it into prime and non-prime channels for dual modular clock hopping. Nodes then perfomr coordinated hopping while exchanging neighbour information via a coordinate-assisted three-way handshake, updating the DNL, INL, and IDN lists. Prior to any transmission, nodes sense the Primary Radio activity to avoid occupied channels. Termination occurs only when all $N-1$ neighbours have been confirmed, and all IDN nodes have been directly verified, guaranteeing a fully validated network topology.

\begin{figure}[htbp]
	\centering
	\includegraphics[width=0.85\linewidth]{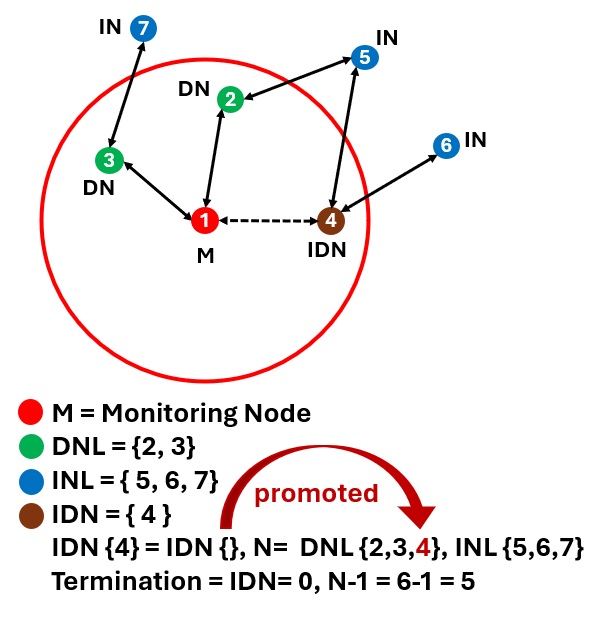} 
	\caption{Controlled termination }
	\label{fig:NlistaN}
\end{figure}

\begin{algorithm}[htbp]
	\smaller 
	\caption{\mbox{MR-Dual Modular Clock Algorithm}}
	\DontPrintSemicolon
	Input: $N$, total number of nodes \& Coordinates \;
	Observed $m_i$, available channels \;
	Divide $m_i$ set into prime ($M_p$) and non-prime ($N_p$) sets \;
	Initialize index $j1_i^{0}$ [0,$m_i$)	randomly for first half \;
	Initialize index $j2_i^{0}$ [0,$m_i$)  randomly for second half \;
	\While {node i not achieve  IDN=0 \& N-1 with all nodes} {
		choose $R1$ from [1, $m_i$) randomly for first half \;
		choose $R2$ from [1, $m_i$) randomly for second half \;			
		\For { {$t_i$= 0} to $m_i$ }{
			\textbf{{start first half of a timeslot}\;}
			$j1_{\text{i}}^{t_i+1}$ = ($j1_{\text{i}}^{t_i}$ + $R1$) mod ($m_i$) \;
			
			\eIf {$M_p$ $>$ 0} { \textbf{c1} = $c_{\text{Mp, ($j1_{i}^{t_i+1}$) mod ($M_p$)}}$ // hop on prime }{c1 = $c_{\text{i, $j1_{i}^{t_i+1}$}}$ // hop on $m_i$ if $M_p=0$\;} 
			\enspace \textbf {sense} the PR Activity on $c1$ \;	 
			
			\enspace attempt a rendezvous only when $c1$ is not occupied\;
			\enspace \If {channel (c1) is matched}  {process handshake}  
			\enspace \While {coordinate-assisted handshake} {
				\textbf {Classify the neighbour list}\;
				\enspace \eIf { directly handshake} {list in DNL}
				{list in INL or IDN as per coordinates}
				\enspace \textbf  {Termination in first half}\;		
				\enspace \eIf {node i achieved IDN=0 \& N-1} {terminate rendezvous}
				{node i wait to complete first half of timeslot\;} \textbf{end while}}
			\textbf{{start second half of a timeslot}\;}		
			$j2_{\text{i}}^{t_i+1}$ = ($j2_{\text{i}}^{t_i}$ + R2) mod ($m_i$) \;				
			\eIf {$N_p$ $>$ 0} {	
				\textbf{c2} = $c_{\text{Np, ($j2_{i}^{t_i+1}$) mod ($N_p$)}}$ // hop on non-prime
			}{ 
				c2 = $c_{\text{i, $j2_{i}^{t_i+1}$}}$ // hop on $m_i$ if $N_p=0$}
			\enspace \If {$(c2 = c1 )$ // only possible if $M_p$ or $N_p$ = 0\;} {    
				($j2_{\text{i}}^{t_i+1}$ + 1) mod ($m_i$) \;
				c2 = $c_{\text{i, $j2_{i}^{t_i+1}$}}$ }
			\enspace \textbf{sense} the PR Activity on $c2$ \; 
			\enspace attempt a rendezvous only when $c2$ is not occupied\;
			\enspace \If {channel (c2) is matched}  { process handshake}  
			\enspace \While {coordinate-assisted handshake} { 
				 \textbf {Classify the neighbour list}\;
			\enspace \eIf { directly handshake} {list in DNL}
			{list in INL or IDN as per coordinates}
			\enspace \textbf  {Termination in second half}\;		
			\enspace \eIf {node i achieved IDN=0 \& N-1} {terminate rendezvous}
			{node i wait to complete second half of timeslot\;} \textbf{end while}}
			\textbf {$t_i$ =  $t_i + 1$ \;}	
			\textbf{end for}}
		\textbf{end while}}
\end{algorithm}

\section{Topology validation model}

To evaluate the correctness of neighbour discovery, the discovered network topology is compared with the actual ground topology derived from node coordinates. The validation measures how accurately the rendezvous protocol reconstructs the true network connectivity.

\subsection{Ground Topology Model}
We assume a set of N static cognitive radio nodes randomly deployed in a two-dimensional area. Each node i is represented by its coordinates $x_i = (x_i, y_i)$ and operates with a homogeneous transmission range r. In this work, the transmission space is fixed as r=100m. 
The network connectivity is modelled using the Unit Disk Model (UDM), where two nodes are considered directly connected if their Euclidean distance is within the transmission range\cite{Clark1990,Gilbert1961,Penrose2003,Kuhn2008}. The ground topology is defined as: 
\begin{equation}
	GT=(N,E^\star)
\end{equation}
where $E^\star$ represents the set of actual communication links between nodes. A link exists between nodes i and j, if: 
\begin{equation}
	\{i,j\}\in E^\star \iff \|x_i-x_j\|_2\le r,\qquad
	GT=(N,E^\star)
\end{equation}
Based on this connectivity, the ground direct neighbour list of node $i$ is defined as:
\begin{equation}
	\label{eq:DNLstar}
	DNL_i^\star=\{\,j\in N\setminus\{i\}:\ \|x_i-x_j\|_2\le r\,\}
\end{equation}
Nodes not directly connected but reachable through multihop paths form the ground indirect neighbour list, defined as:
\begin{equation}
	\label{eq:INLstar}
	INL_i^\star=\{\,j\in N\setminus(\{i\}\cup DNL_i^\star):\ i \rightsquigarrow_{GT} j\,\}.
\end{equation}
where $i$ $\rightsquigarrow_{GT}$ $j$ indicates that node $j$ is reachable from node $i$ through multihop paths in the ground topology.

\subsection{Discovered Topology Model}
During protocol execution, nodes discover neighbours through rendezvous and handshake exchanges. The discovered topology is defined as:
\begin{equation}
	DT=(N,E)
\end{equation}
where $E$ represents the set of links established through successful handshakes.
The discovered direct neighbour list of node $i$ is therefore;
\begin{equation}
	\label{eq:DNL}
	DNL_i=\{\,j\in N\setminus\{i\}:\ \{i,j\}\in E\,\},
\end{equation}
Similarly, nodes that are learned through multihop neighbour information form the discovered indirect neighbour list;

\begin{equation}
	\label{eq:INL}
	INL_i=\{\,j\in N\setminus(\{i\}\cup DNL_i):\ i \rightsquigarrow_{DT} j\,\}.
\end{equation}
These neighbour lists represent the topology constructed by the rendezvous protocol.
\subsection{Topology Match Metric}
To quantify the correctness of the discovered topology, the direct neighbour lists of the discovered topology are compared with the ground topology. Since indirect neighbours depend on multihop information exchange, the validation focuses on direct neighbours.

The per-node topology match (PTM) is defined as the ratio of correctly discovered direct neighbours to the actual direct neighbours.

\begin{equation}
	\mathrm{PTM}_i = \bigg(\dfrac{DNL_i \cap DNL_i^{\star}}{DNL_i^{\star}}\bigg) \times 100 
\end{equation}
This metric represents the percentage of correctly discovered direct neighbours for node i.

The Complete Topology Match (CTM) of the network is then obtained by averaging the PTM values across all nodes:
\begin{equation}
	\label{eq:CTM}
	\text{CTM}=\sum_{i\in N} \dfrac{PTM_i}{N}
\end{equation}
CTM represents the overall percentage of correctly discovered topology and is used to compute the Average Topology Match (ATM) metric in the performance evaluation.

\section{Performance Evaluation}
The proposed MR-DMCA is evaluated using a simulation framework that 
divides each timeslot into two half-slots to implement the dual clock hopping mechanism. The simulation environment models a post-disaster cognitive radio network scenario and incorporates MAC layer operations, dual channel selection, multihop rendezvous, a three-way handshake (3-WH) mechanism, random network topology, and PR activity \cite{chigan,6}. For fair comparison, the benchmark protocols RCS, MCA \cite{5} and EMCA \cite{6} are evaluated under the same dual-rendezvous framework 2RATS (Two Rendezvous Attempts in a timeslot).
     
\subsection{Simulation Setup}

The proposed MR-DMCA protocol was implemented and evaluated using the NS-3 network simulator. Randomly distributed network topologies consisting of 3, 10, and 20 cognitive radio (CR) nodes were generated using the static mobility model of the BonnMotion tool \cite{bonnmotion_web}. The nodes were deployed within a  1000 $\times$ 1000$~\text{m}^2$ area to emulate a multihop post-disaster communication environment. Each node operates over either 10 or 20 available channels, with asymmetric channel availability across nodes to capture realistic spectrum heterogeneity. A channel similarity index $\left(m\right)$ of 2 or 5 was used to control the degree of overlapping channels among nodes. A three-way handshake (3-WH) mechanism, adopted from the M-DMCA protocol \cite{Ali2026}, was used to enable reliable multihop neighbour discovery.
Primary radio activity was incorporated to model spectrum occupancy, considering two scenarios: no PR activity (0\%) and high PR activity (85\%). The PR activity rates  ($\lambda_X$ and $\lambda_Y$) were adopted from the parameter settings reported in the M-DMCA study.

To obtain statistically reliable results, each simulation scenario was executed 1000 times, and the average values were reported. 
\subsection{Simulation parameters and performance metrics}
The performance of the proposed MR-DMCA protocol is evaluated using three key metrics:

\subsubsection{Average Time to Rendezvous (ATTR)}
The Average Time to Rendezvous (ATTR) is defined as the mean TTR value, first averaged across all nodes (N) in a single simulation run, and then averaged over all simulation runs (R), as given in Eq.~\eqref{eq:ADT}. This metric quantifies the efficiency of the protocol in establishing network connectivity, where lower ATTR indicates faster rendezvous and more efficient neighbour discovery. 

\begin{equation}
	\ ATTR =  \dfrac{1}{R} \sum_{j=1}^{R}  \left( \dfrac{1}{N} \sum_{i=1}^{N} {TTR_i } \right)
	\label{eq:ADT}
\end{equation}

\subsubsection{Average Topology Match (ATM)}
To evaluate the probability of premature termination in the traditional approaches versus the proposed MR-DMCA, we measure the ATM. This metric is defined as the mean CTM in a single simulation run, which is then averaged over all simulation runs (R), as given in Eq.\ref{eq:ATM}. ATM represents the percentage similarity between the actual network topology and the discovered topology obtained by the rendezvous protocol. It is calculated by comparing the neighbour relationships in the discovered topology with the ground truth topology. ATM represents the percentage of correctly discovered neighbours relative to the actual network topology. A higher value indicates more accurate neighbour discovery and more complete topology formation, reflecting the reliability of the protocol.
\begin{equation}
	\text{ATM}= \dfrac{1}{R} \sum_{i=1}^{R} {CTM_i}
	\label{eq:ATM}
\end{equation} 
\subsubsection{Post termination discovery delay (PTDD)}
Post-Termination Discovery Delay (PTDD) measures the additional time required to achieve complete neighbour discovery after the nominal N-1 termination condition is satisfied. In multihop rendezvous protocols, partial discovery may occur due to asynchronous rendezvous, channel hopping mismatches, or incomplete handshakes, causing some neighbours to remain incorrectly identified at the termination point.

To quantify this effect, the protocol execution is allowed to continue beyond the N-1 termination point. PTDD is defined as the difference between ATTR measured when 100\% topology discovery is achieved $\left( ATTR_{ATM} \right)$ and the ATTR measured when N-1 condition is satisfied $\left( ATTR_{N-1} \right)$. The PTDD is therefore defined as the additional time required to achieve complete topology discovery after termination. 
\begin{equation}
	\ PTDD =    {ATTR_{\left(ATM\right)} - ATTR_{\left(N-1\right)} }  %
	\label{eq:PTDD}
\end{equation}

\subsection{Simulation Scenarios}

To analyse the effect of termination conditions on neighbour discovery and topology correctness, two simulation scenarios are considered.
\subsubsection{Baseline Termination Scenario}

In this scenario, traditional rendezvous protocols, including RCS, MCA, EMCA and M-DMCA, follow the conventional N-1 termination condition, where the rendezvous process ends once each node reports N-1 neighbours. In contrast, the proposed MR-DMCA protocol employs a controlled termination mechanism that incorporates coordinate-based validation through the Intended Direct Neighbour (IDN) phase. As a result, MR-DMCA terminates only when both conditions N-1 and IDN=0 (i.e., all intended neighbours have been directly verified) are satisfied. This scenario highlights the premature termination problem present in conventional approaches. 

\subsubsection{Controlled Termination Scenario}
In this scenario, all protocols, including RCS, MCA, EMCA, and the extended M-DMCA (i.e., MR-DMCA) are evaluated using the same termination condition, where nodes terminate the rendezvous process only when both N-1 and IDN=0 conditions are satisfied. This scenario enables a fair comparison by eliminating premature termination and evaluating rendezvous efficiency under enforced complete topology discovery.

\subsection{Average Time to Rendezvous (ATTR)}
\begin{figure}[ht]
	\centering	
	\begin{minipage}[b]{0.5\columnwidth}
		\centering
		\begin{tikzpicture}
			\node[draw, line width=0.5pt, inner sep=2pt] {%
				\includegraphics[width=1.4in]{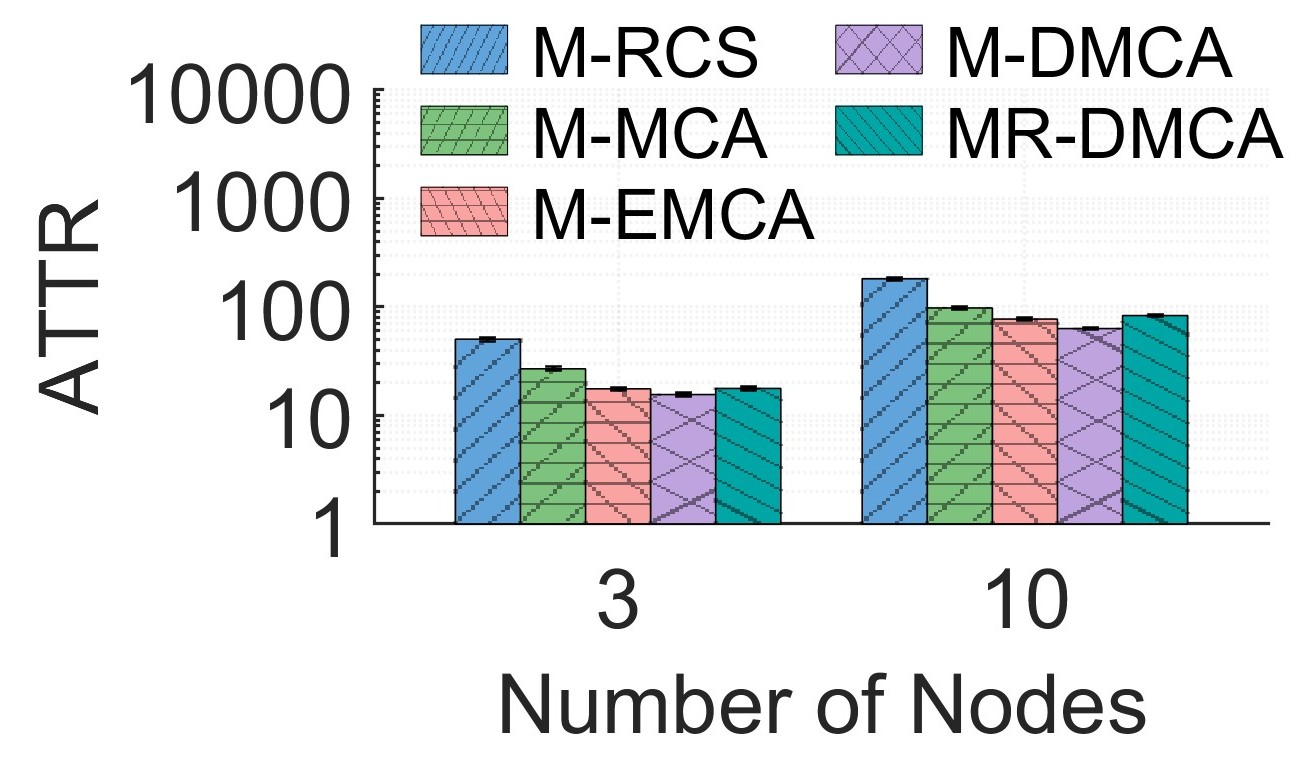}
			};
		\end{tikzpicture}\\[-0.25em]
		{\footnotesize (a)}
	\end{minipage}%
	\begin{minipage}[b]{0.5\columnwidth}
		\centering
		\begin{tikzpicture}
			\node[draw, line width=0.5pt, inner sep=2pt] {%
				\includegraphics[width=1.4in]{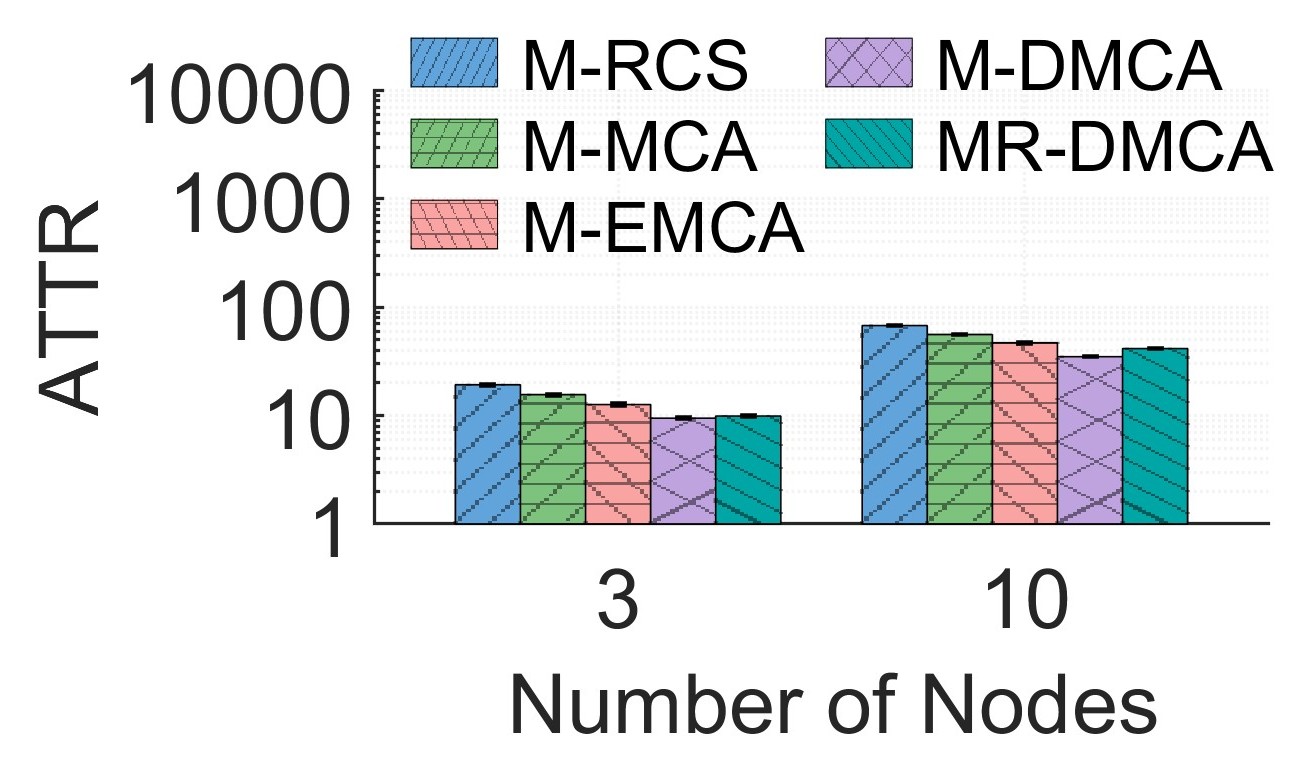}%
			};
		\end{tikzpicture}\\[-0.25em]
		{\footnotesize (b)}
	\end{minipage}
	\caption{Asym 10-CH (a) m=2 (b) m=5  with 0\%PR}\hfill
	\label{fig:C_R3bb_R5aa}
\end{figure}

\begin{figure}[ht]
	\centering	
	\begin{minipage}[b]{0.5\columnwidth}
		\centering
		\begin{tikzpicture}
			\node[draw, line width=0.5pt, inner sep=2pt] {%
				\includegraphics[width=1.4in]{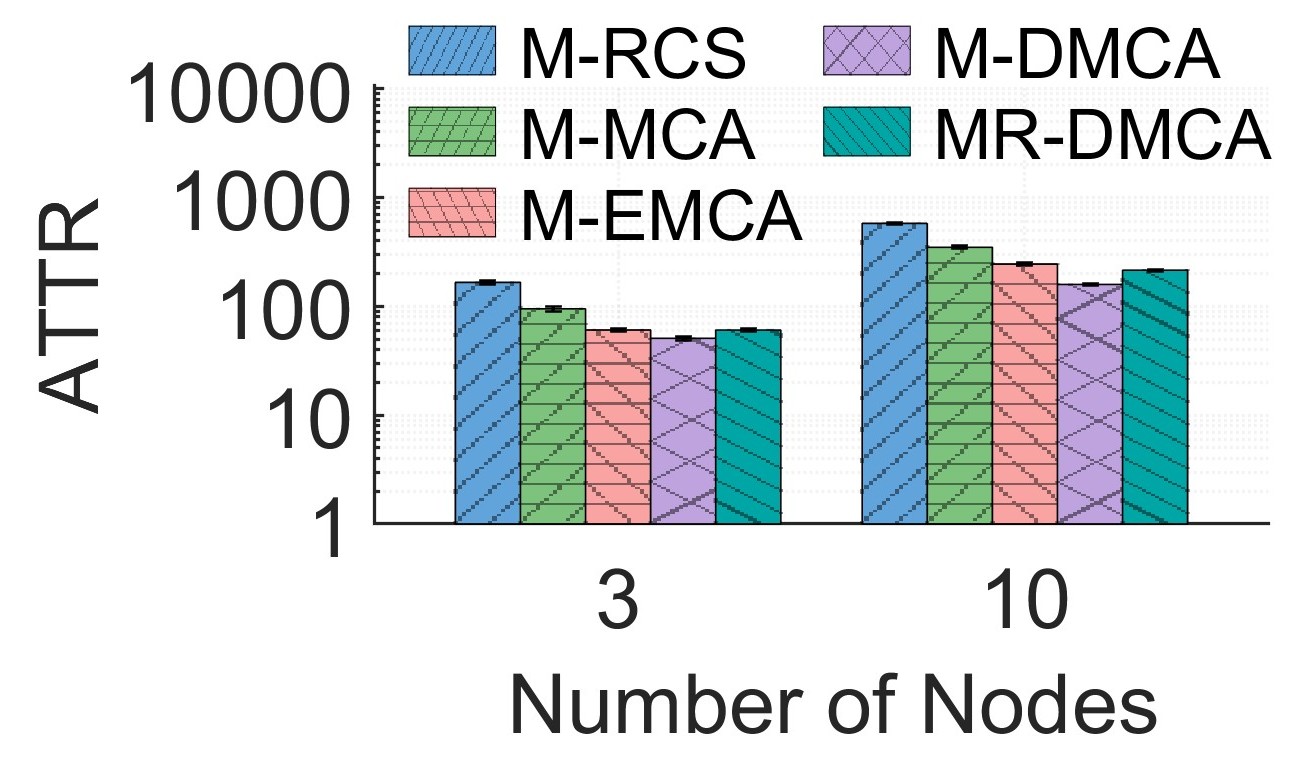}
			};
		\end{tikzpicture}\\[-0.25em]
		{\footnotesize (a)}
	\end{minipage}%
	\begin{minipage}[b]{0.5\columnwidth}
		\centering
		\begin{tikzpicture}
			\node[draw, line width=0.5pt, inner sep=2pt] {%
				\includegraphics[width=1.4in]{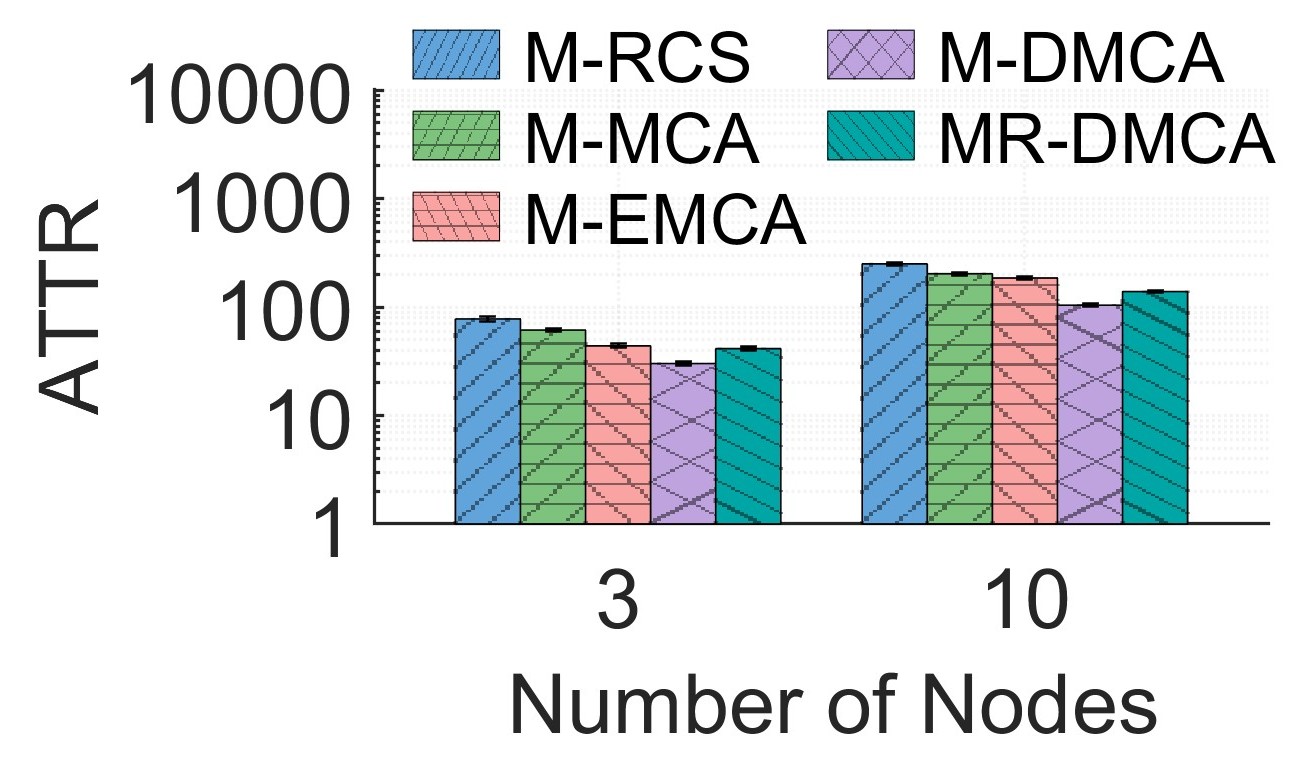}%
			};
		\end{tikzpicture}\\[-0.25em]
		{\footnotesize (b)}
	\end{minipage}
	\caption{Asym 10-CH (a) m=2 (b) m=5 with 85\%PR}\hfill
	\label{fig:C_R4bb_R5bb}
\end{figure}

Fig.~\ref{fig:C_R3bb_R5aa} presents the ATTR performance for the baseline scenario under 0\% PR activity with asymmetric channel sets (10 channels). The results show that the ATTR of MR-DMCA is slightly higher than EMCA and M-DMCA but remains lower than RCS and MCA. This behaviour occurs because conventional protocols terminate the rendezvous process upon satisfying the N-1 condition,
while MR-DMCA continues until the additional IDN = 0 condition is met, ensuring that all neighbours are correctly discovered. Similar trends are observed for both channel similarity indices m=2 and m=5 across all network sizes. On average, MR-DMCA compromises ATTR by approximately 14\% compared to the baseline M-DMCA protocol.
When PR activity increases to 85\%, the ATTR increases for all protocols due to the reduced availability of idle channels, as illustrated in Fig.~\ref{fig:C_R4bb_R5bb}. However, the overall trend shows: MR-DMCA outperforms RCS, MCA, and EMCA, while M-DMCA exhibits slightly lower ATTR due to its earlier termination based solely on the N-1 condition.

\begin{figure}[ht]
	\centering	
	\begin{minipage}[b]{0.5\columnwidth}
		\centering
		\begin{tikzpicture}
			\node[draw, line width=0.5pt, inner sep=2pt] {%
				\includegraphics[width=1.4in]{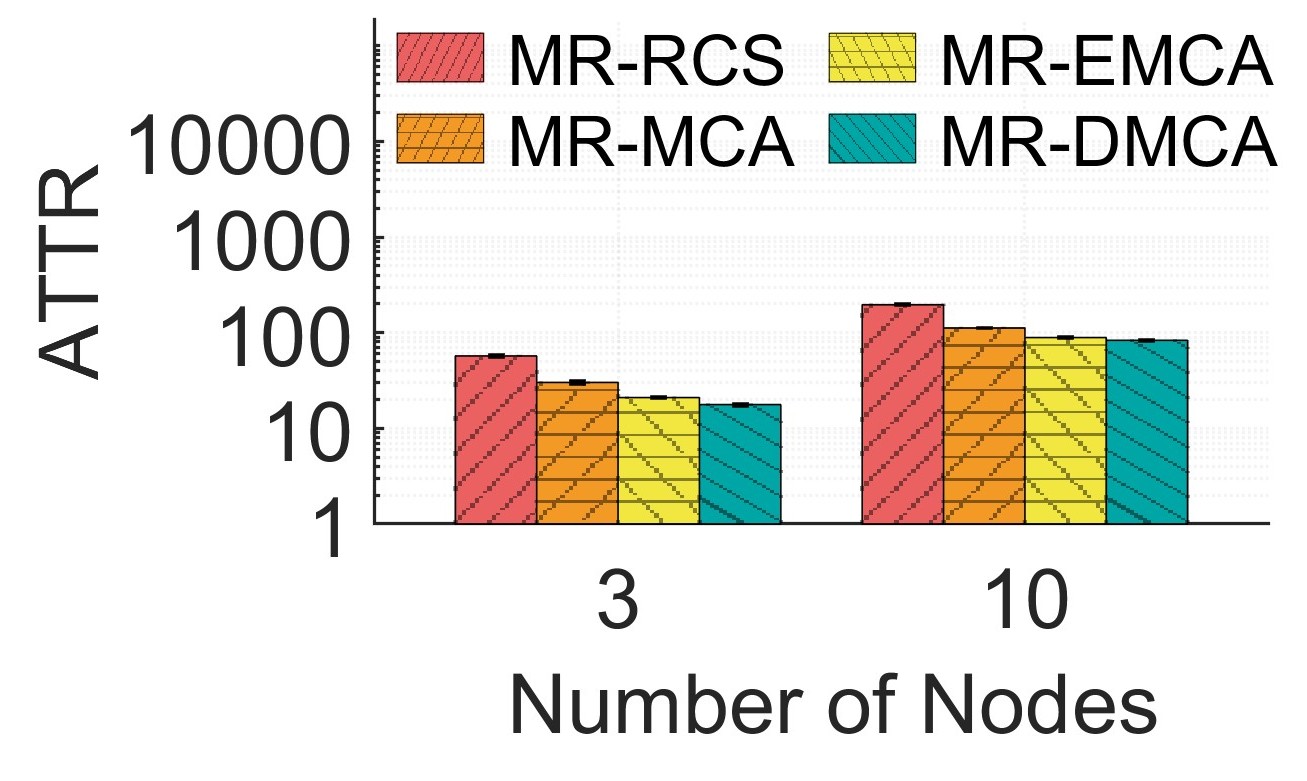}
			};
		\end{tikzpicture}\\[-0.25em]
		{\footnotesize (a)}
	\end{minipage}%
	\begin{minipage}[b]{0.5\columnwidth}
		\centering
		\begin{tikzpicture}
			\node[draw, line width=0.5pt, inner sep=2pt] {%
				\includegraphics[width=1.4in]{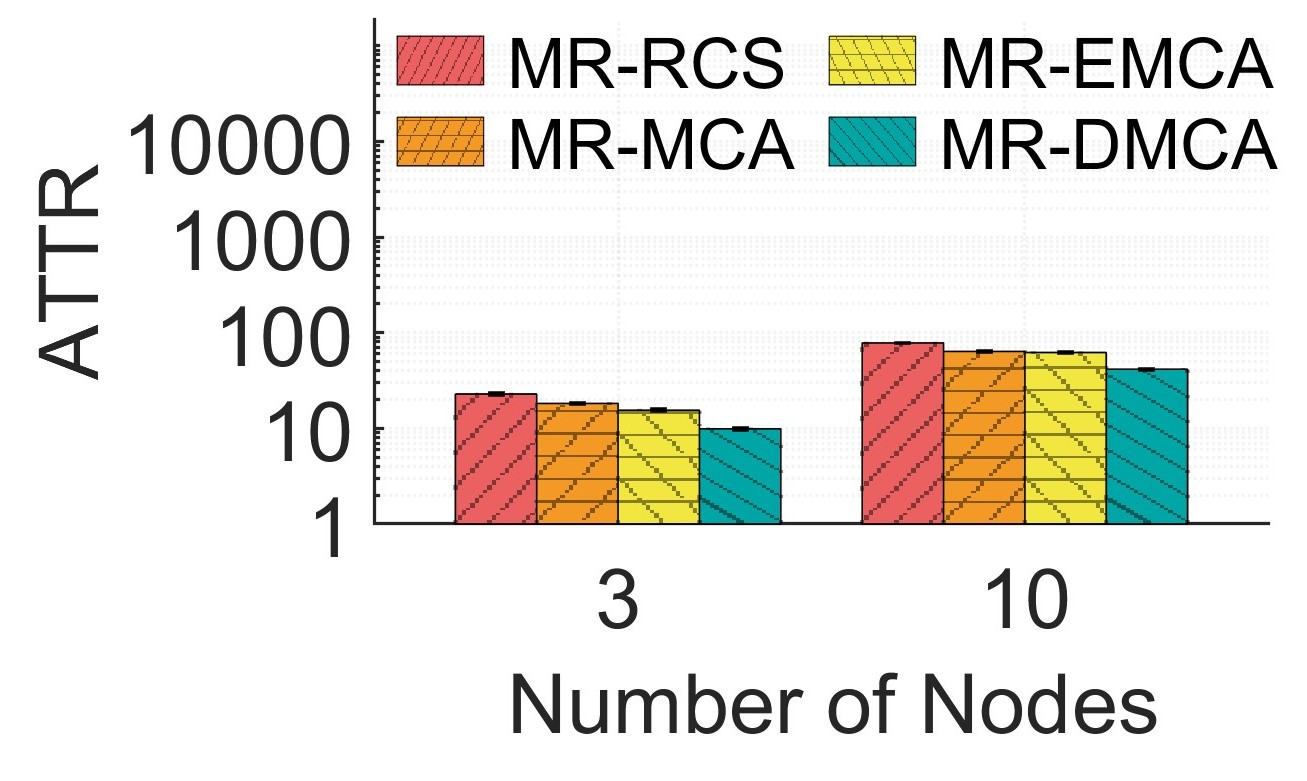}%
			};
		\end{tikzpicture}\\[-0.25em]
		{\footnotesize (b)}
	\end{minipage}
	\caption{Asym 10-CH (a) m=2 (b) m=5 with 0\%PR}\hfill
	\label{fig:IDN_R3bb_R5aa}
\end{figure}

\begin{figure}[ht]
	\centering	
	\begin{minipage}[b]{0.5\columnwidth}
		\centering
		\begin{tikzpicture}
			\node[draw, line width=0.5pt, inner sep=2pt] {%
				\includegraphics[width=1.4in]{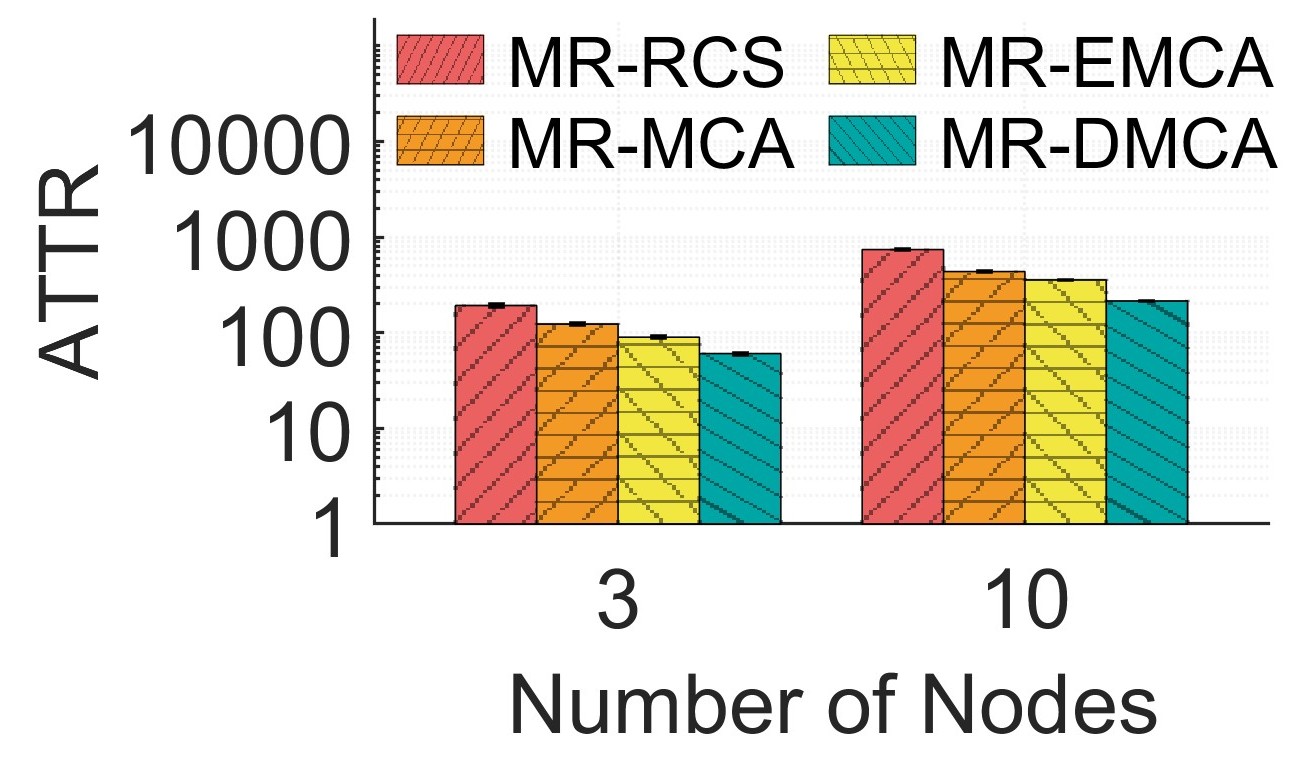}
			};
		\end{tikzpicture}\\[-0.25em]
		{\footnotesize (a)}
	\end{minipage}%
	\begin{minipage}[b]{0.5\columnwidth}
		\centering
		\begin{tikzpicture}
			\node[draw, line width=0.5pt, inner sep=2pt] {%
				\includegraphics[width=1.4in]{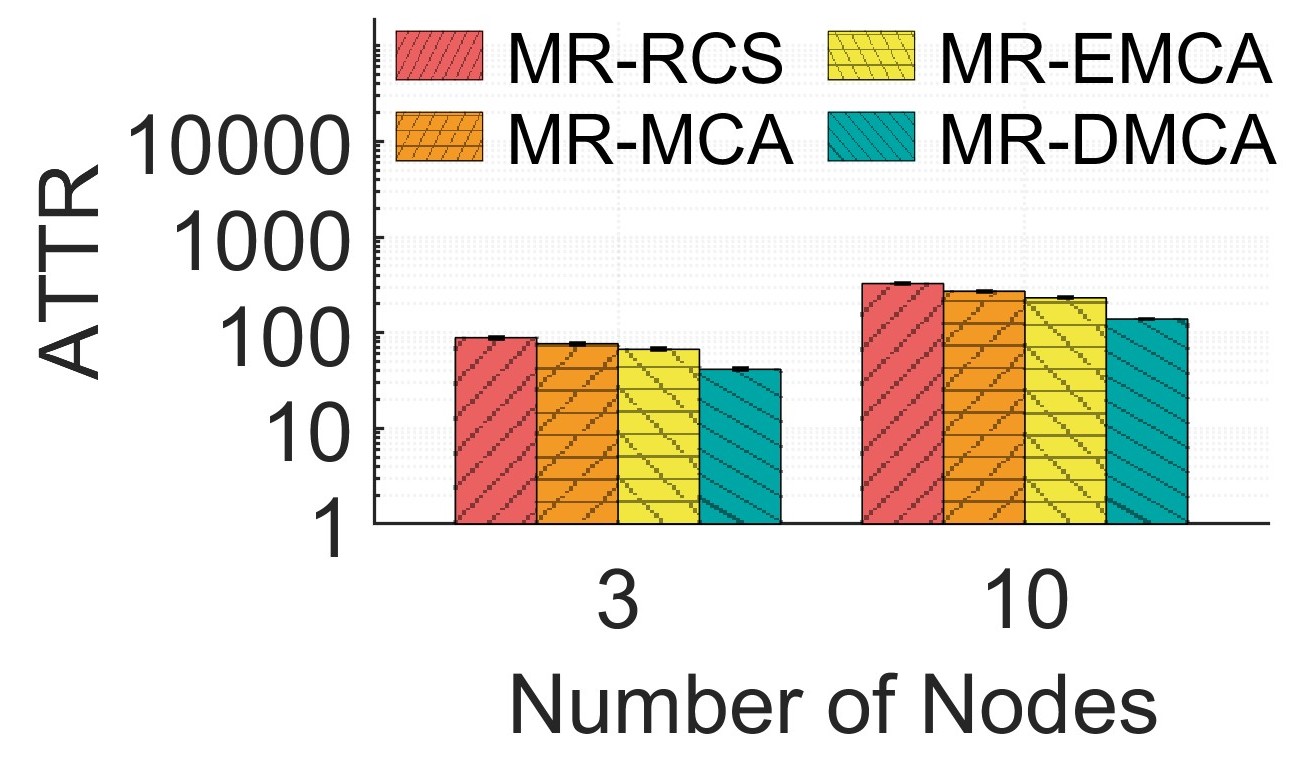}%
			};
		\end{tikzpicture}\\[-0.25em]
		{\footnotesize (b)}
	\end{minipage}
	\caption{Asym 10-CH (a) m=2 (b) m=5  with 85\%PR}\hfill
	\label{fig:IDN_R4bb_R5bb}
\end{figure}

To further evaluate the effect of termination conditions, simulations were conducted under the controlled termination scenario, where all protocols employ both N-1 and IDN = 0 conditions before terminating. As shown in Fig.~\ref{fig:IDN_R3bb_R5aa} and Fig.~\ref{fig:IDN_R4bb_R5bb}, MR-DMCA achieves lower ATTR compared to the other protocols under this setting. Since all protocols now wait until IDN=0, they continue the rendezvous process beyond the N-1 condition, which increases their overall discovery time. In contrast, MR-DMCA is specifically designed to operate under this termination mechanism and therefore maintains better rendezvous efficiency. Similar trends
under both 0\% and 85\% PR activity (Figures ~\ref{fig:IDN_R3bb_R5aa} and \ref{fig:IDN_R4bb_R5bb}).

Furthermore, increasing the channel similarity index to m=5 reduces ATTR for all protocols, as the higher probability of channel overlap facilitates faster rendezvous, while the relative performance trend among protocols remains consistent. Overall, the results indicate that although MR-DMCA may exhibit slightly higher ATTR in the baseline scenario due to its stricter termination requirement, it prevents premature termination and ensures reliable neighbour discovery. When all protocols operate under the same termination conditions in the controlled scenario, MR-DMCA consistently achieves lower rendezvous delay, demonstrating its efficiency in achieving correct and complete topology discovery.

\subsection{Average Topology Match (ATM)}
\begin{figure}[ht]
	\centering	
	\begin{minipage}[b]{0.5\columnwidth}
		\centering
		\begin{tikzpicture}
			\node[draw, line width=0.5pt, inner sep=2pt] {%
				\includegraphics[width=1.4in]{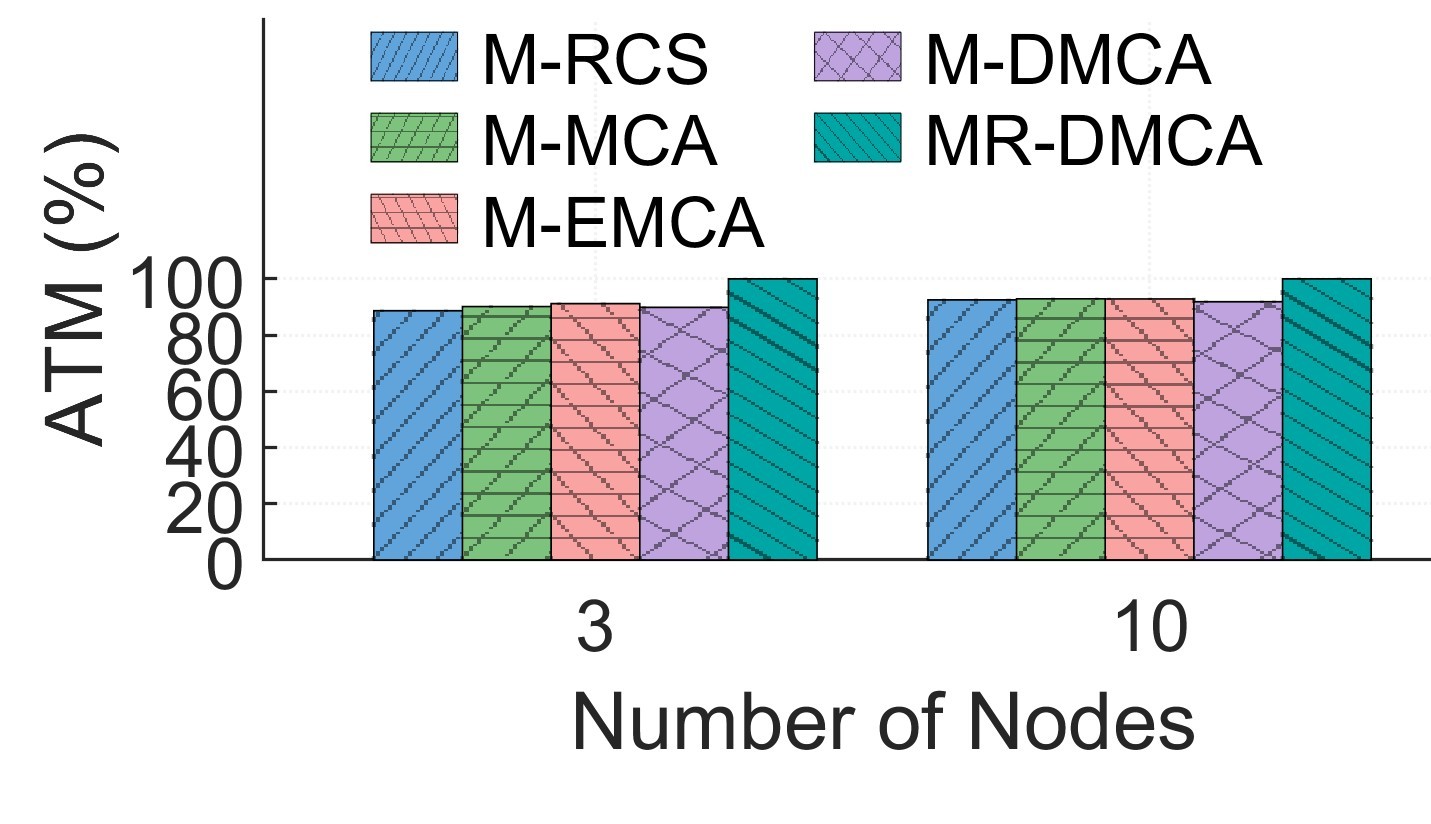}
			};
		\end{tikzpicture}\\[-0.25em]
		{\footnotesize (a)}
	\end{minipage}%
	\begin{minipage}[b]{0.5\columnwidth}
		\centering
		\begin{tikzpicture}
			\node[draw, line width=0.5pt, inner sep=2pt] {%
				\includegraphics[width=1.4in]{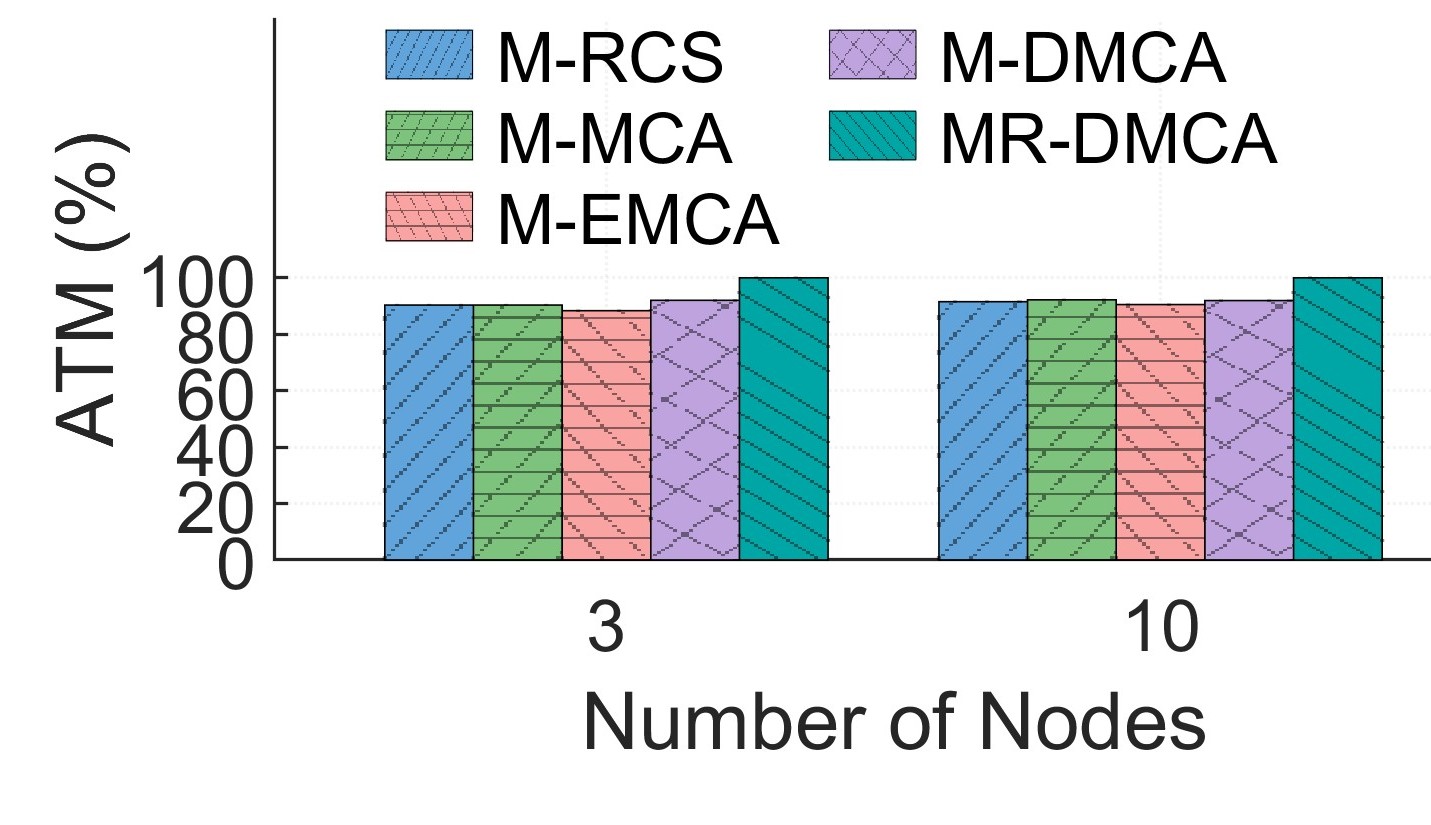}%
			};
		\end{tikzpicture}\\[-0.25em]
		{\footnotesize (b)}
	\end{minipage}
	\caption{ATM for Asym 10-CH (a) m=2 (b) m=5 with 0\%PR}\hfill
	\label{fig:TM_R3bb_R5aa}
\end{figure}

\begin{figure}[ht]
	\centering	
	\begin{minipage}[b]{0.5\columnwidth}
		\centering
		\begin{tikzpicture}
			\node[draw, line width=0.5pt, inner sep=2pt] {%
				\includegraphics[width=1.4in]{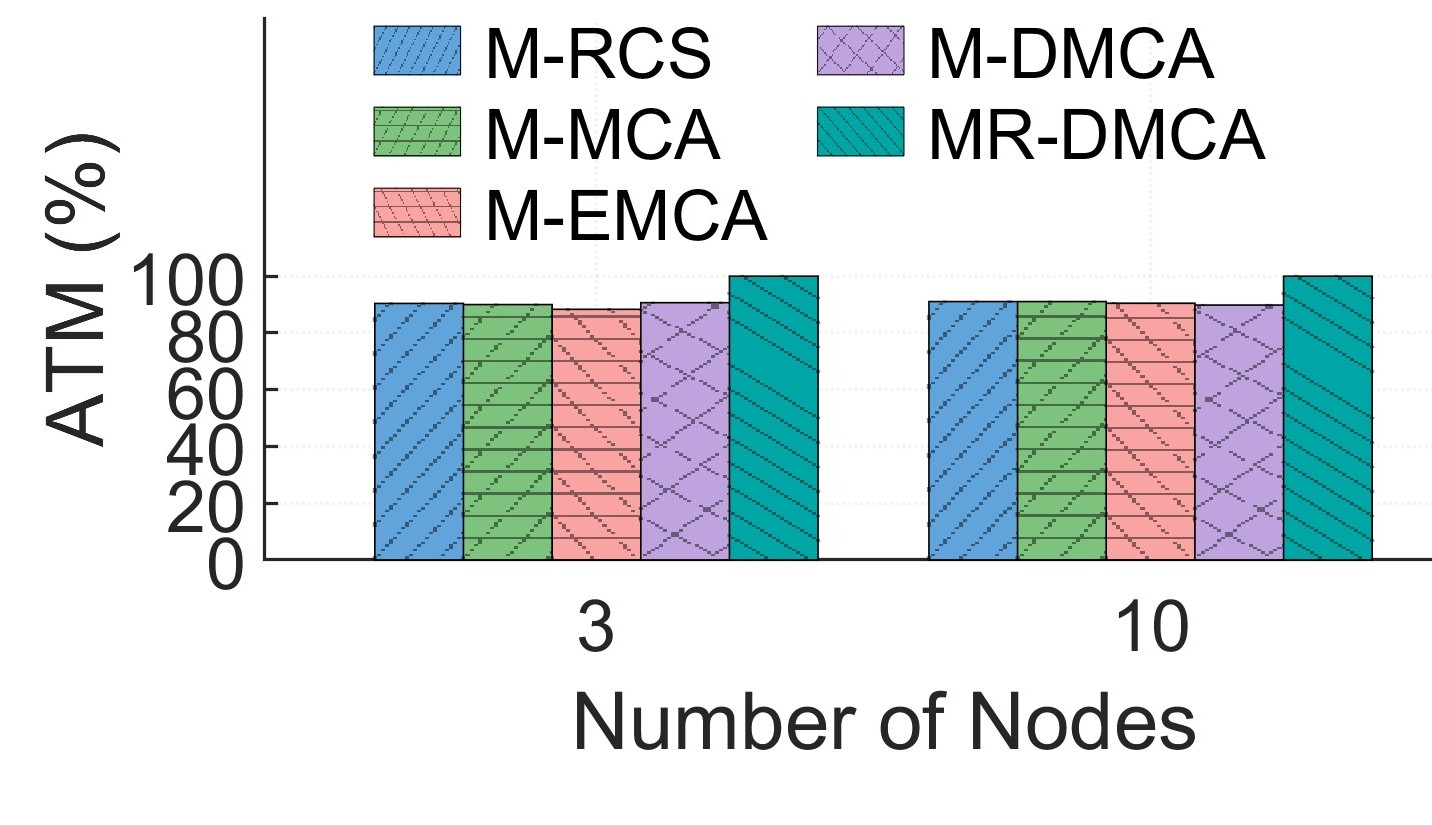}
			};
		\end{tikzpicture}\\[-0.25em]
		{\footnotesize (a)}
	\end{minipage}%
	\begin{minipage}[b]{0.5\columnwidth}
		\centering
		\begin{tikzpicture}
			\node[draw, line width=0.5pt, inner sep=2pt] {%
				\includegraphics[width=1.4in]{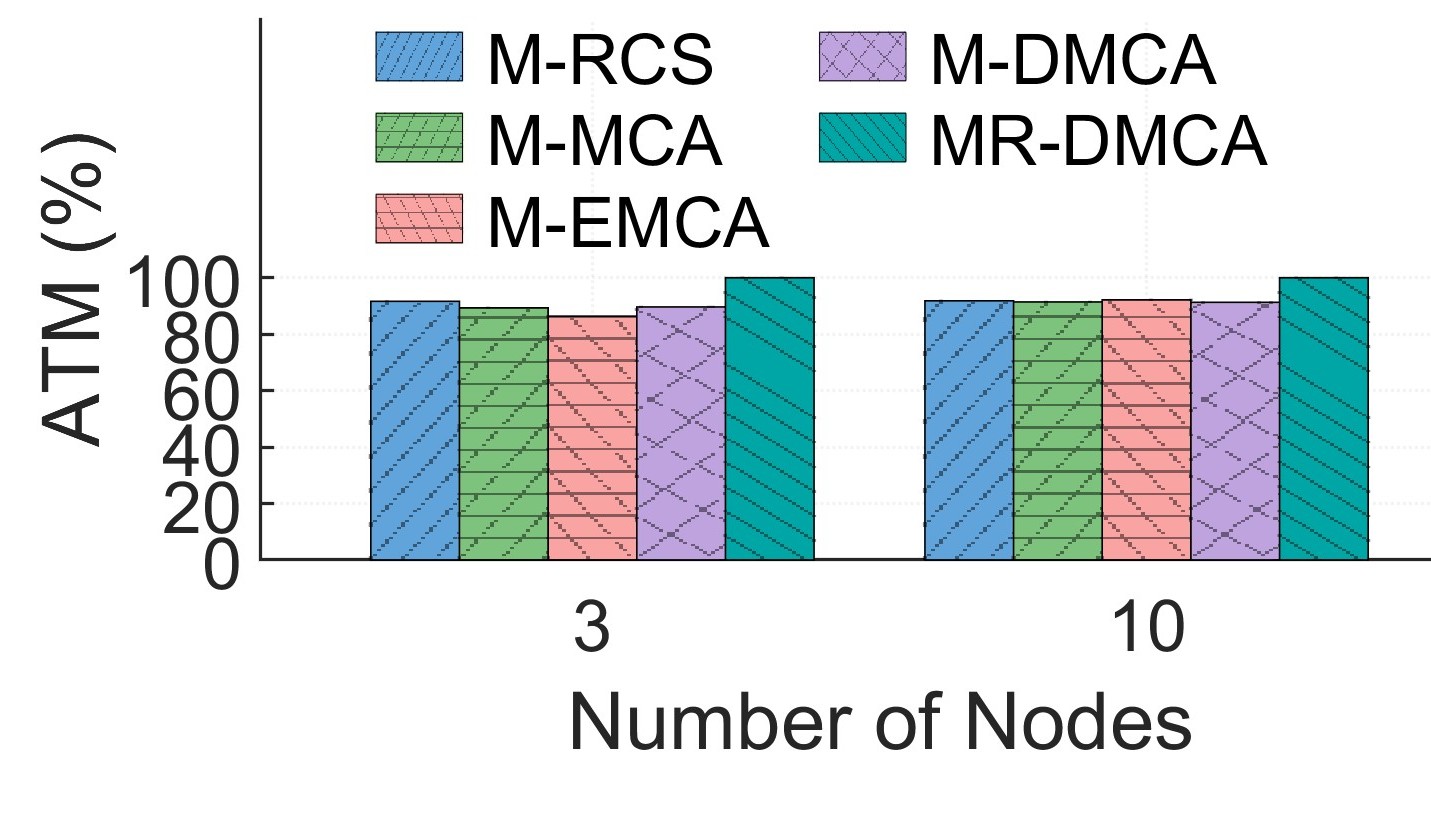}%
			};
		\end{tikzpicture}\\[-0.25em]
		{\footnotesize (b)}
	\end{minipage}
	\caption{ATM for Asym 10-CH (a) m=2 (b) m=5 with 85\%PR}\hfill
	\label{fig:TM_R4bb_R5bb}
\end{figure}

ATM represents the percentage of correctly discovered neighbours compared to the actual network topology. Fig.~\ref{fig:TM_R3bb_R5aa} and \ref{fig:TM_R4bb_R5bb} present the ATM results for the baseline termination scenario corresponding to the ATTR results discussed earlier. The results show that traditional protocols, including RCS, MCA, EMCA, and M-DMCA, achieve ATM values ranging between approximately 88\% and 90\% under both PR activity conditions. This indicates that these protocols may terminate prematurely when the N-1 condition is satisfied, even though some neighbours have not yet been correctly confirmed. As a result, a portion of the network topology remains incorrectly discovered.
In contrast, the proposed MR-DMCA protocol consistently achieves 100\% ATM in both PR activity scenarios. This improvement is achieved because MR-DMCA introduces the additional IDN=0 validation condition, which ensures that nodes do not terminate the rendezvous process until all direct neighbours within communication range are correctly verified. Consequently, MR-DMCA eliminates premature termination and guarantees complete and accurate topology discovery.

\begin{figure}[ht]
	\centering	
	\begin{minipage}[b]{0.5\columnwidth}
		\centering
		\begin{tikzpicture}
			\node[draw, line width=0.5pt, inner sep=2pt] {%
				\includegraphics[width=1.4in]{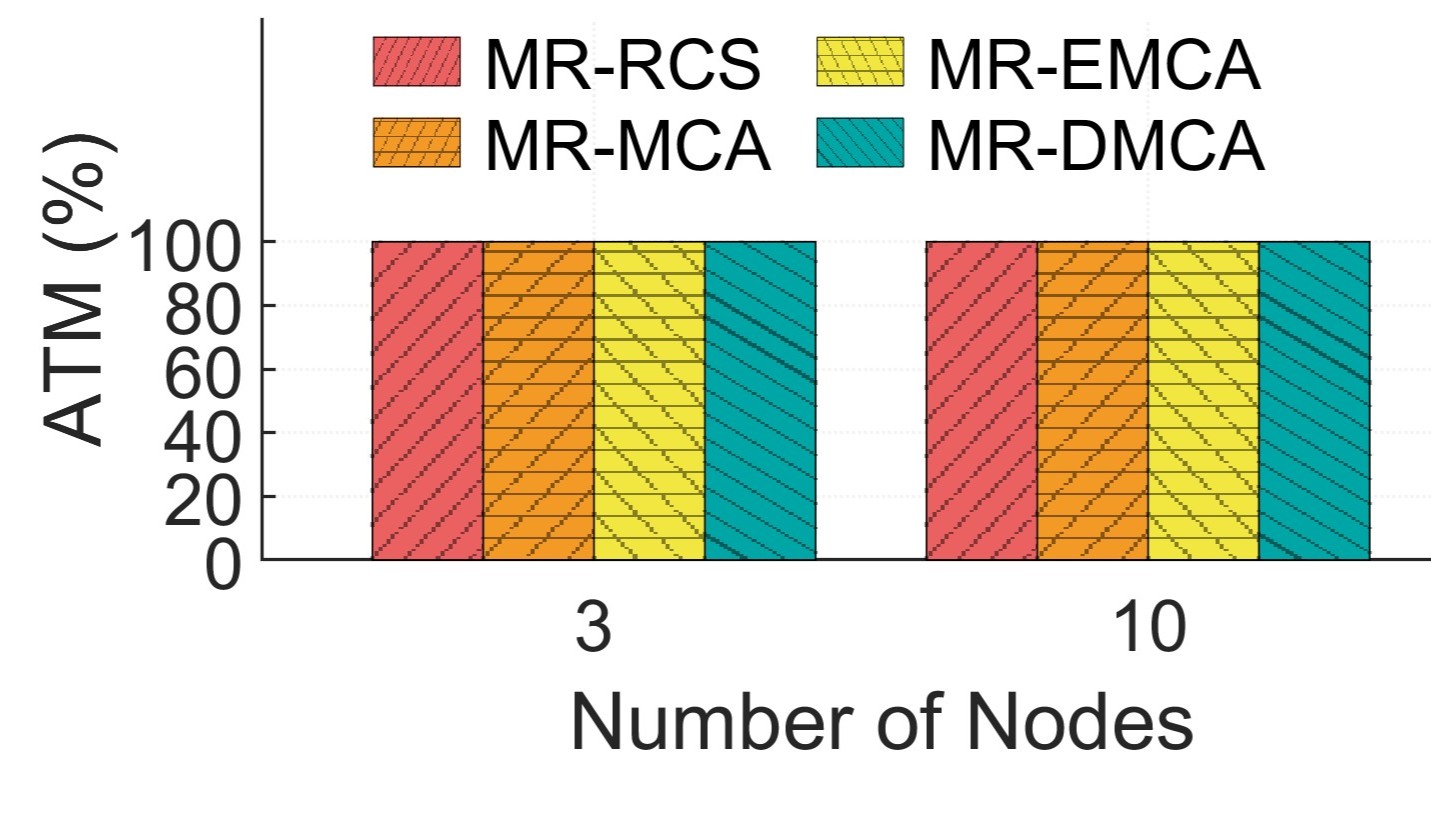}
			};
		\end{tikzpicture}\\[-0.25em]
		{\footnotesize (a)}
	\end{minipage}%
	\begin{minipage}[b]{0.5\columnwidth}
		\centering
		\begin{tikzpicture}
			\node[draw, line width=0.5pt, inner sep=2pt] {%
				\includegraphics[width=1.4in]{IDN_TM_R3bbN.jpg}%
			};
		\end{tikzpicture}\\[-0.25em]
		{\footnotesize (b)}
	\end{minipage}
	\caption{ATM for Asym 10-CH (a) m=2 (b) m=5 with 0\%PR }\hfill
	\label{fig:IDN_TM_R3bb_R5aa}
\end{figure}

\begin{figure}[ht]
	\centering	
	\begin{minipage}[b]{0.5\columnwidth}
		\centering
		\begin{tikzpicture}
			\node[draw, line width=0.5pt, inner sep=2pt] {%
				\includegraphics[width=1.4in]{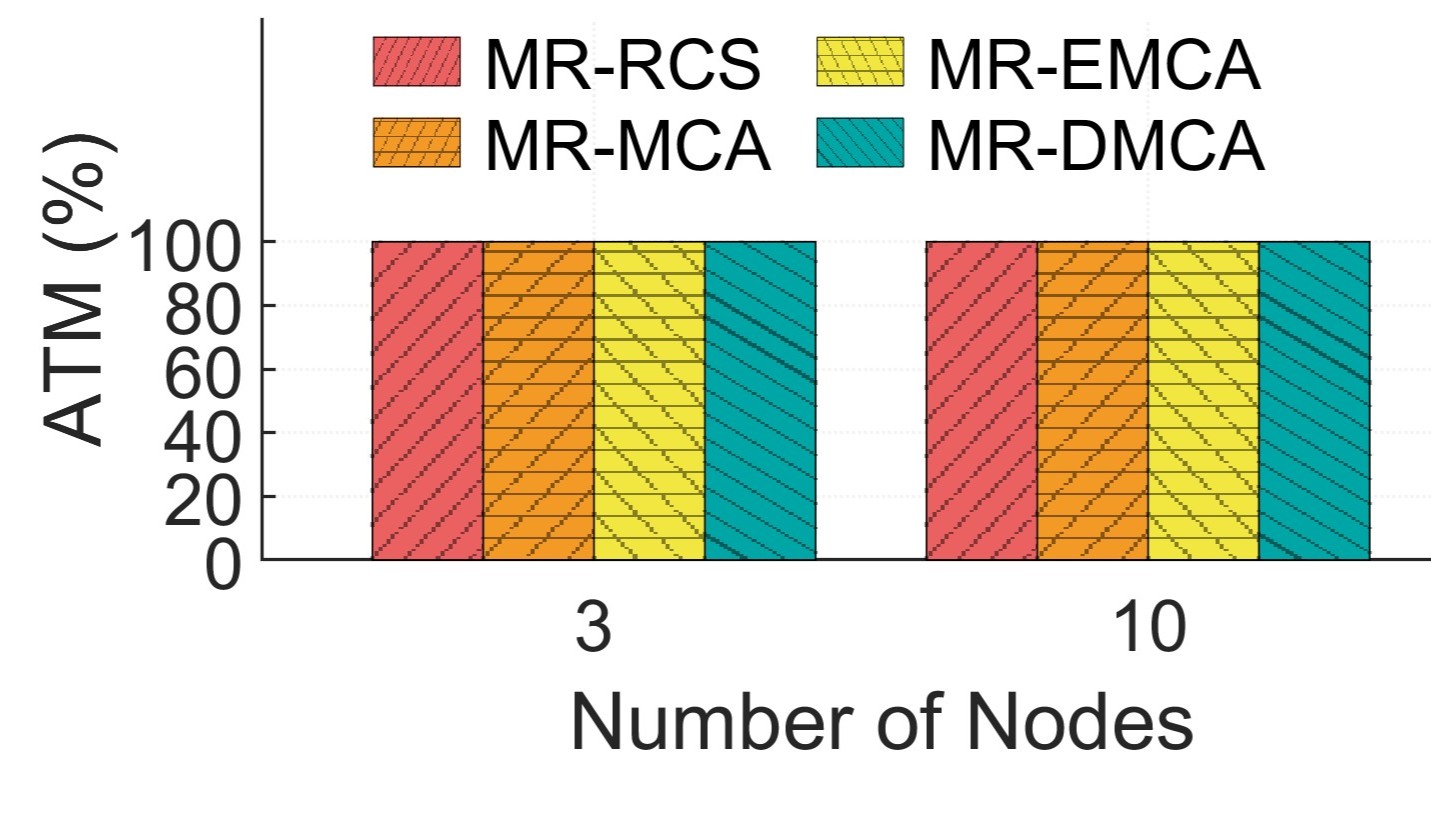}
			};
		\end{tikzpicture}\\[-0.25em]
		{\footnotesize (a)}
	\end{minipage}%
	\begin{minipage}[b]{0.5\columnwidth}
		\centering
		\begin{tikzpicture}
			\node[draw, line width=0.5pt, inner sep=2pt] {%
				\includegraphics[width=1.4in]{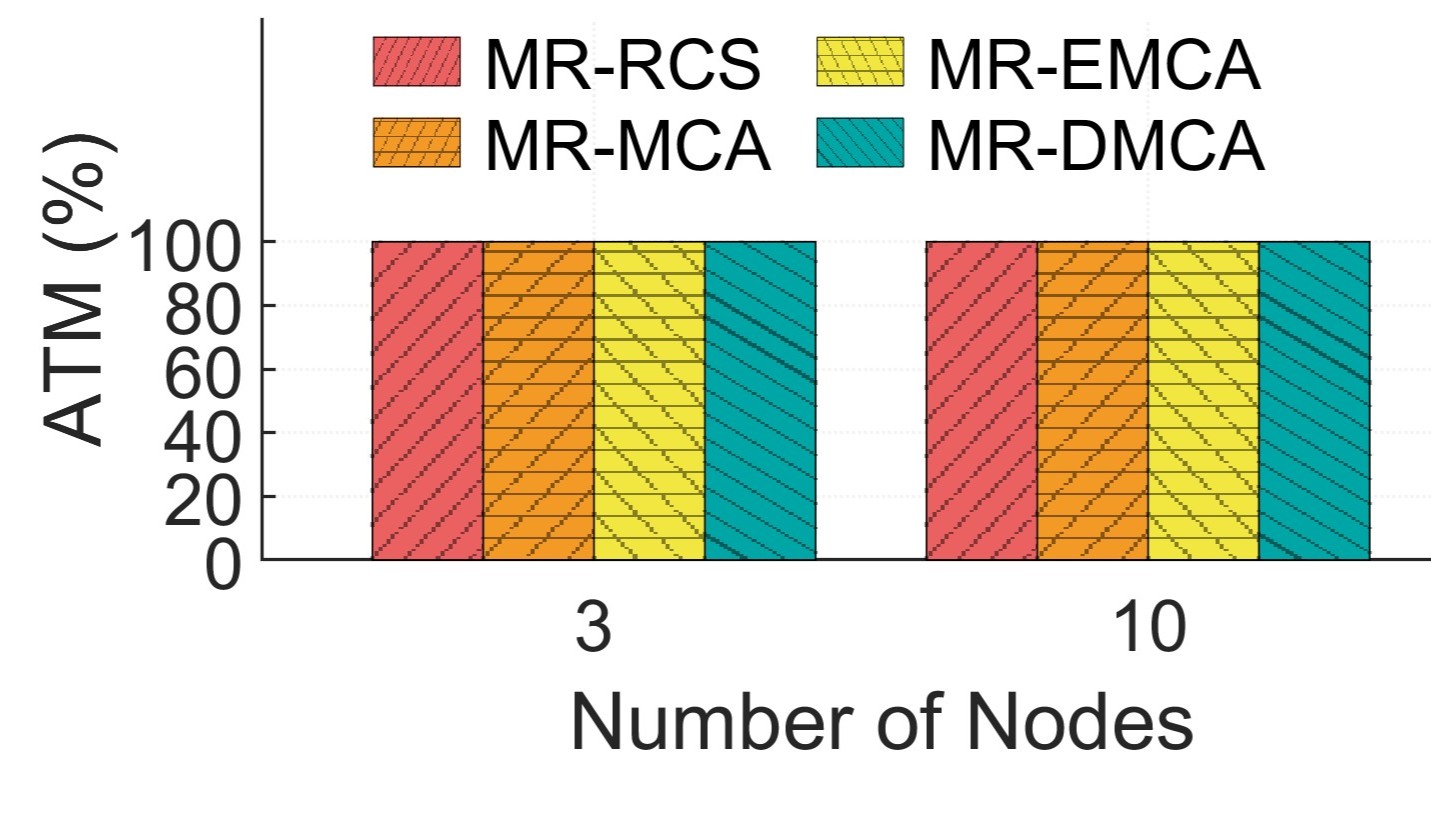}%
			};
		\end{tikzpicture}\\[-0.25em]
		{\footnotesize (b)}
	\end{minipage}
	\caption{ATM for Asym 10-CH (a) m=2 (b) m=5 with 85\%PR }\hfill
	\label{fig:IDN_TM_R4bb_R5aa}
\end{figure}

To further examine the effect of termination conditions, the ATM results are also evaluated under the controlled termination scenario, where all protocols employ both the N-1 and IDN=0 termination conditions. As shown in Fig.~\ref{fig:IDN_TM_R3bb_R5aa} and \ref{fig:IDN_TM_R4bb_R5aa}, all protocols achieve 100\% ATM under this setting. Since the additional validation phase forces nodes to continue the discovery process until all direct neighbours are confirmed, the resulting topology becomes fully accurate for all protocols.
Overall, these results highlight that while conventional protocols may achieve faster termination under the baseline scenario, they may suffer from incomplete topology discovery. In contrast, MR-DMCA ensures reliable neighbour discovery and correct topology formation without relying on external termination control.

\subsection{Post Termination Discovery delay (PTDD)}
\begin{figure}[ht]
	\centering	
	\begin{minipage}[b]{0.5\columnwidth}
		\centering
		\begin{tikzpicture}
			\node[draw, line width=0.5pt, inner sep=2pt] {%
				\includegraphics[width=1.4in]{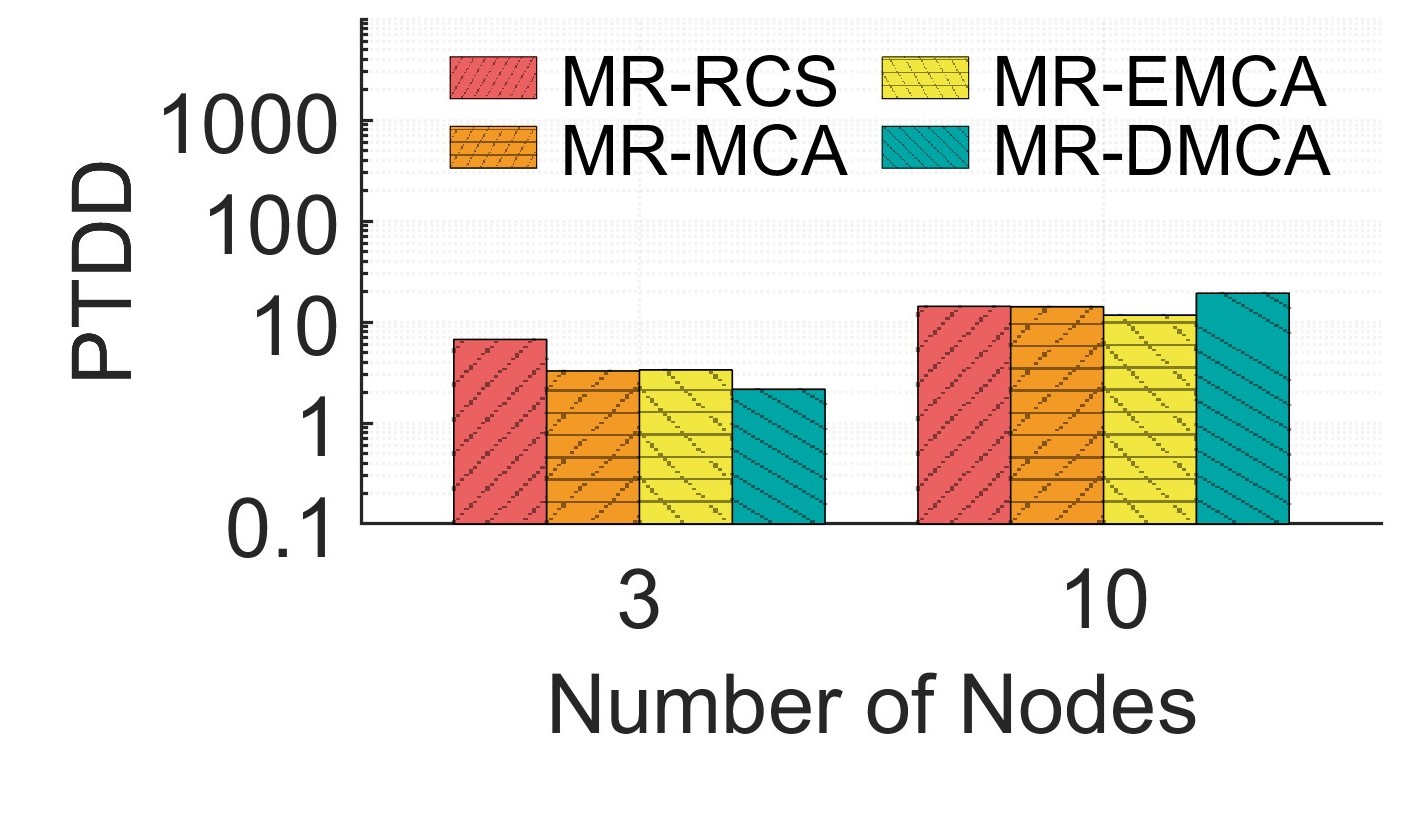}
			};
		\end{tikzpicture}\\[-0.25em]
		{\footnotesize (a)}
	\end{minipage}%
	\begin{minipage}[b]{0.5\columnwidth}
		\centering
		\begin{tikzpicture}
			\node[draw, line width=0.5pt, inner sep=2pt] {%
				\includegraphics[width=1.4in]{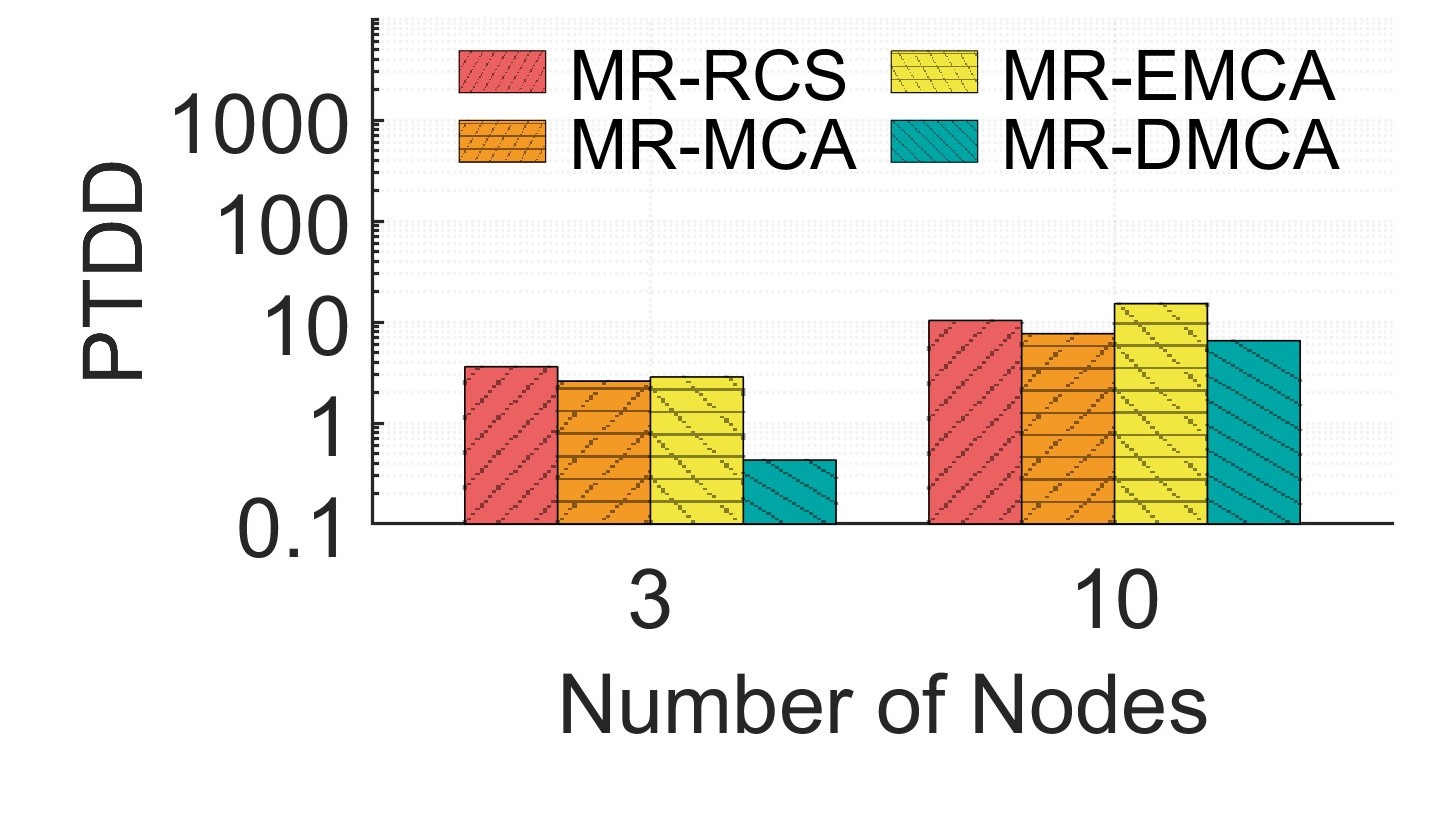}%
			};
		\end{tikzpicture}\\[-0.25em]
		{\footnotesize (b)}
	\end{minipage}
	\caption{PTDD Asym 10-CH (a) m=2 (b) m=5 with 0\%PR}\hfill
	\label{fig:PTDD_R3bb_R5aa}
\end{figure}

\begin{figure}[ht]
	\centering	
	\begin{minipage}[b]{0.5\columnwidth}
		\centering
		\begin{tikzpicture}
			\node[draw, line width=0.5pt, inner sep=2pt] {%
				\includegraphics[width=1.4in]{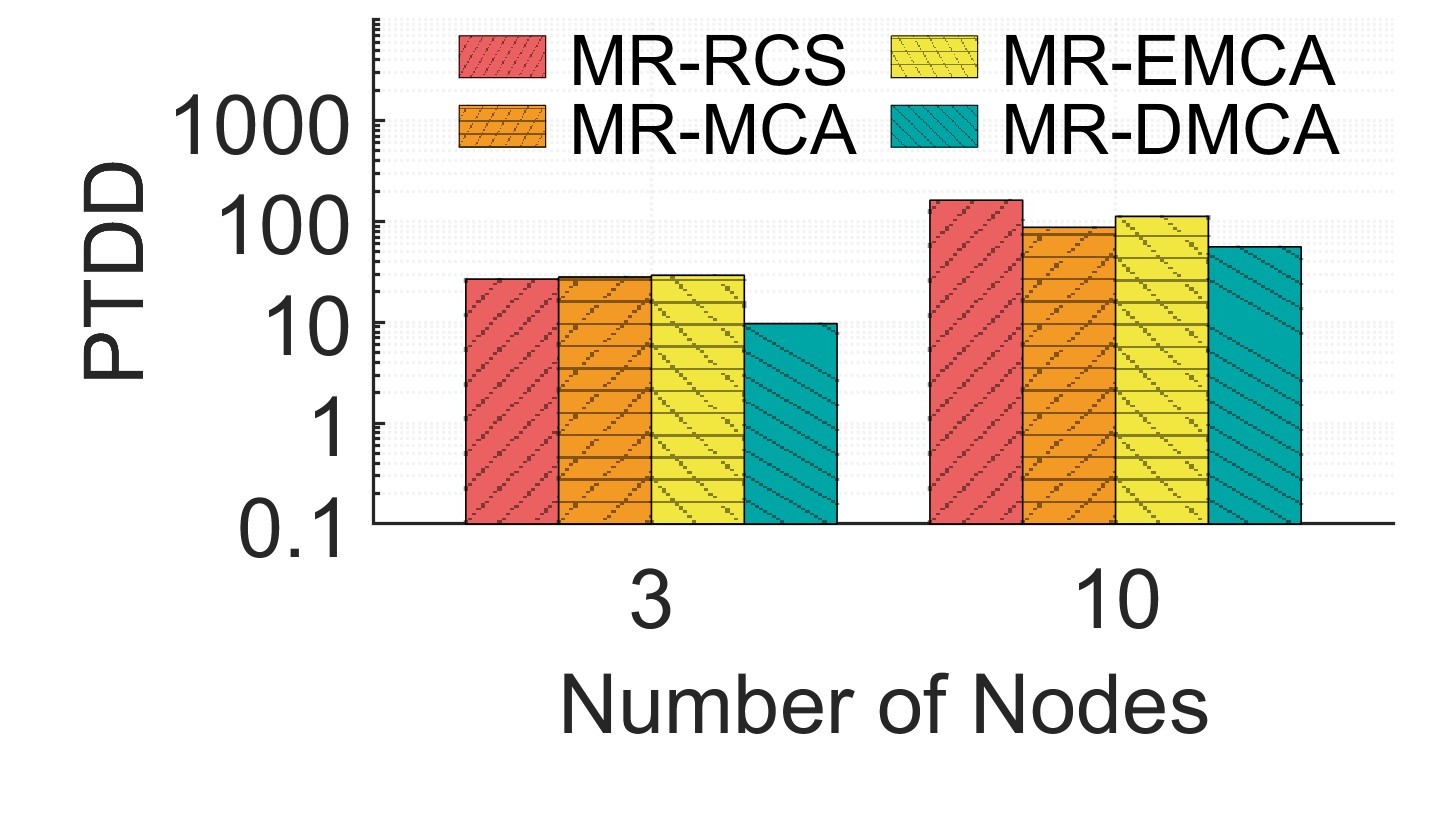}
			};
		\end{tikzpicture}\\[-0.25em]
		{\footnotesize (a)}
	\end{minipage}%
	\begin{minipage}[b]{0.5\columnwidth}
		\centering
		\begin{tikzpicture}
			\node[draw, line width=0.5pt, inner sep=2pt] {%
				\includegraphics[width=1.4in]{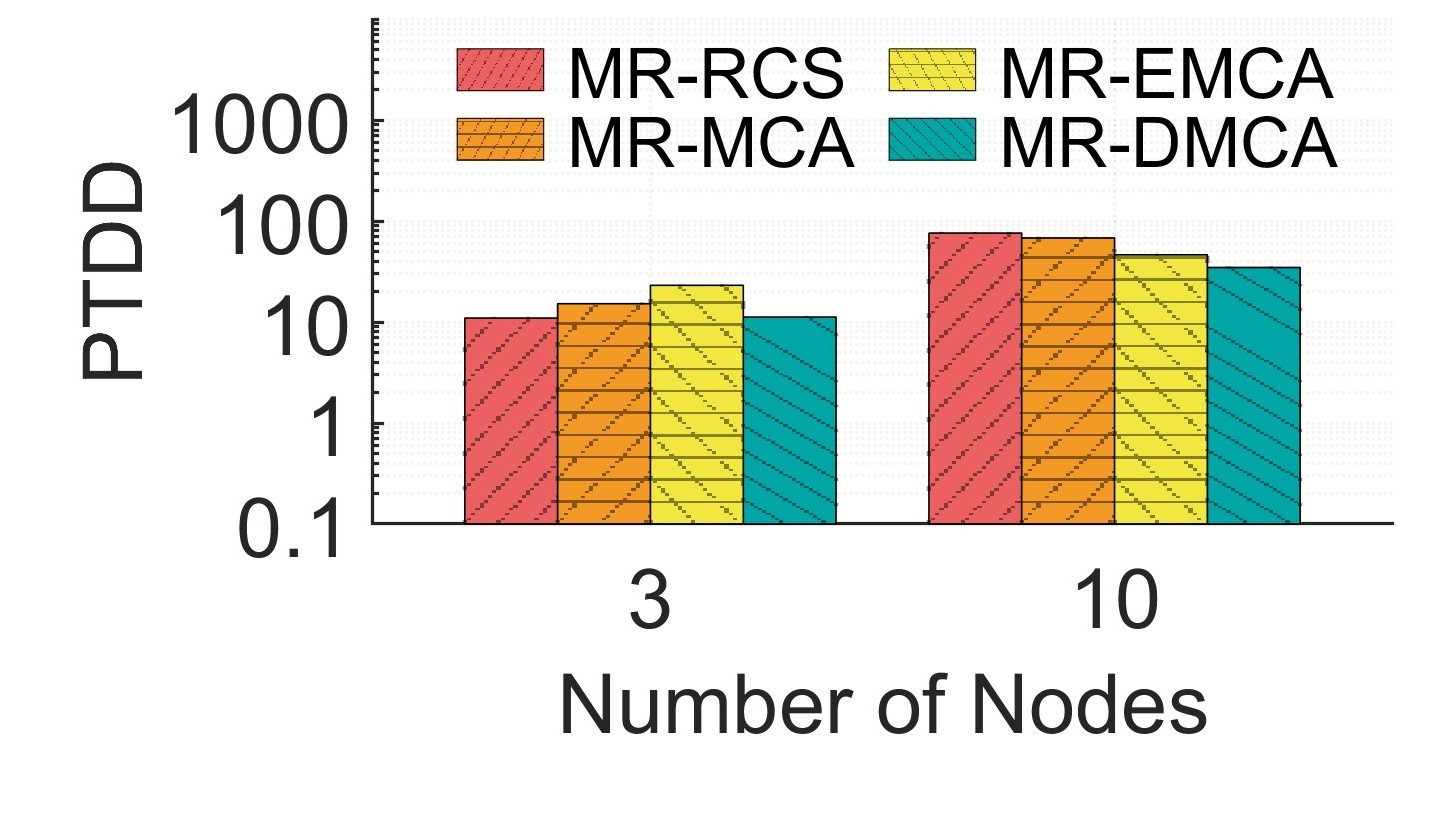}%
			};
		\end{tikzpicture}\\[-0.25em]
		{\footnotesize (b)}
	\end{minipage}
	\caption{PTDD for Asym 10-CH (a) m=2 (b) m=5 with 85\%PR}\hfill
	\label{fig:PTDD_R4bb_R5bb}
\end{figure}

PTDD represents the additional time required to achieve complete topology discovery, calculated as the difference between ATTR under the traditional N-1 termination condition and ATTR when discovery continues until full topology correctness. The Conventional protocols show lower ATTR because they terminate once the N-1 condition is met, even if some direct neighbours remain unverified. In contrast, reliable discovery mechanisms require the rendezvous process to continue until all direct neighbours are confirmed, which introduces additional discovery time.
Fig.~\ref{fig:PTDD_R3bb_R5aa} illustrates the PTDD results for the case of m=2 with 0\% PR activity. The results show that achieving complete topology discovery requires approximately 3 and 14 timeslots of additional time for networks with 3 and 10 nodes, respectively. Similarly, under higher PR activity (85\%), the PTDD increases due to reduced channel availability, as shown in Fig.~\ref{fig:PTDD_R4bb_R5bb}. In this case, approximately 23 and 104 timeslots of additional discovery time are required for networks with 3 and 10 nodes, respectively.

These results highlight the trade-off between early termination and topology correctness. While conventional approaches may appear faster due to premature termination, the additional time quantified by PTDD represents the delay required to achieve complete and accurate neighbour discovery. By incorporating topology validation during the rendezvous process, MR-DMCA avoids this post-termination discovery delay and ensures reliable topology formation.

\subsection{Network and Channel Scalability Analysis}
\begin{figure}[ht]
	\centering	
	\begin{minipage}[b]{0.5\columnwidth}
		\centering
		\begin{tikzpicture}
			\node[draw, line width=0.5pt, inner sep=2pt] {%
				\includegraphics[width=1.4in]{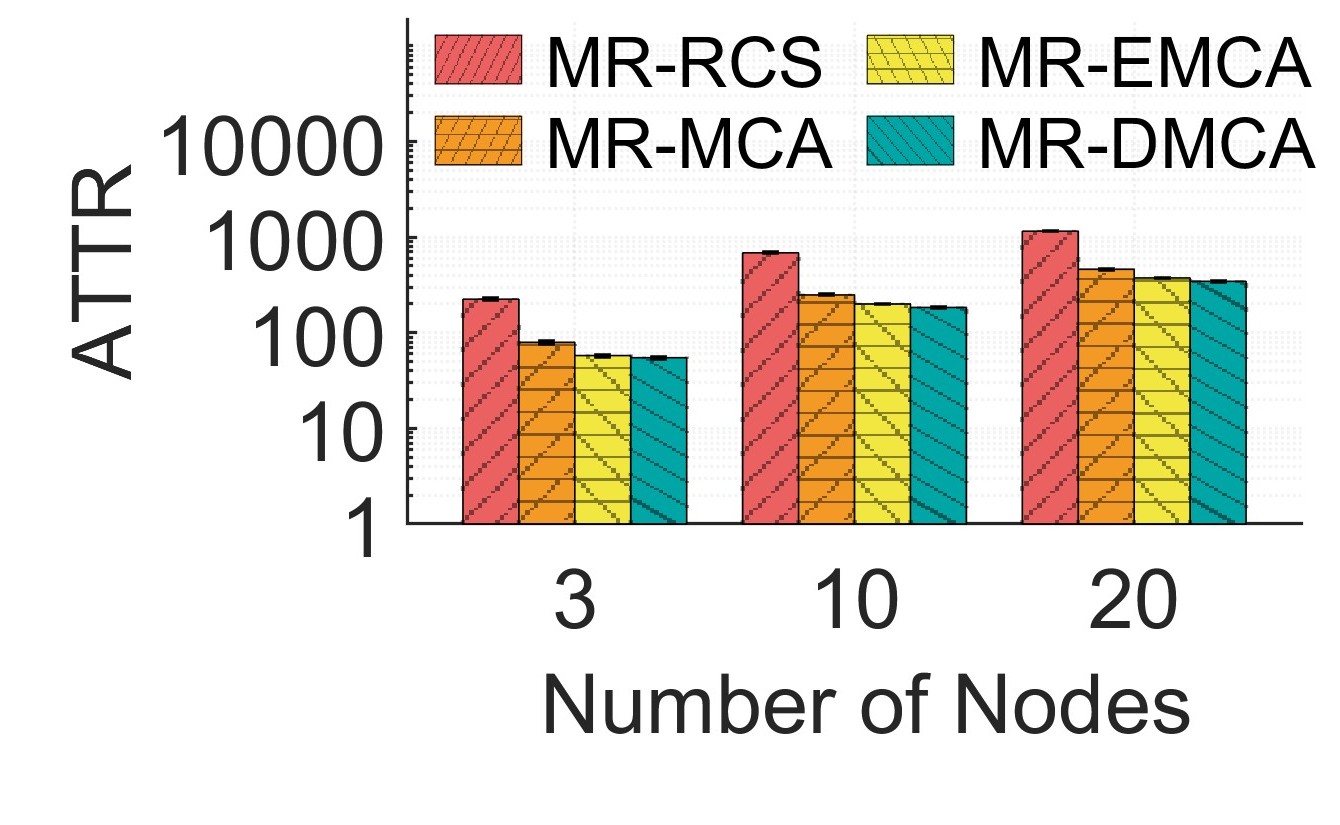}
			};
		\end{tikzpicture}\\[-0.25em]
		{\footnotesize (a)}
	\end{minipage}%
	\begin{minipage}[b]{0.5\columnwidth}
		\centering
		\begin{tikzpicture}
			\node[draw, line width=0.5pt, inner sep=2pt] {%
				\includegraphics[width=1.4in]{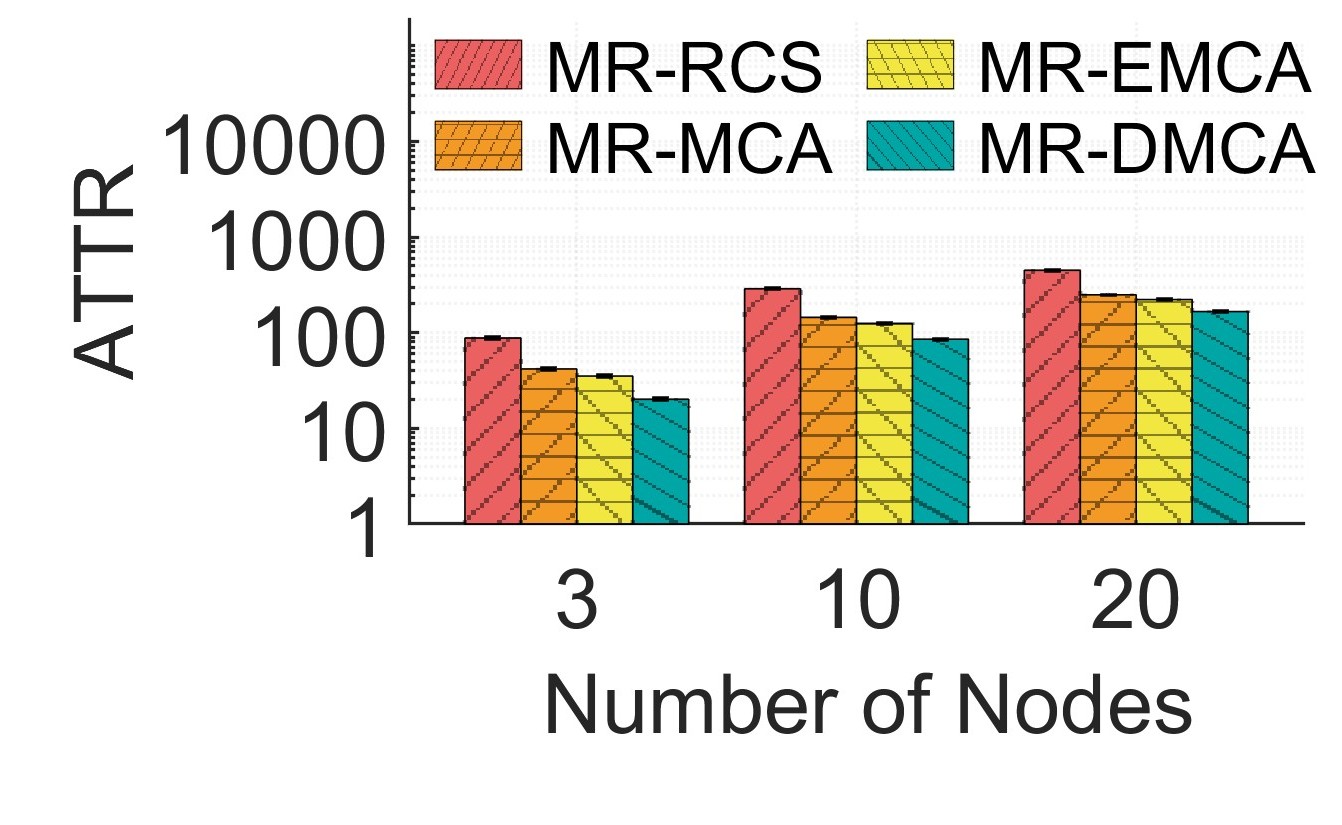}%
			};
		\end{tikzpicture}\\[-0.25em]
		{\footnotesize (b)}
	\end{minipage}
	\caption{Asym 20-CH (a) m=2 (b) m=5 with 0\%PR}\hfill
	\label{fig:IDN_R7aa_R8aa}
\end{figure}

\begin{figure}[ht]
	\centering	
	\begin{minipage}[b]{0.5\columnwidth}
		\centering
		\begin{tikzpicture}
			\node[draw, line width=0.5pt, inner sep=2pt] {%
				\includegraphics[width=1.4in]{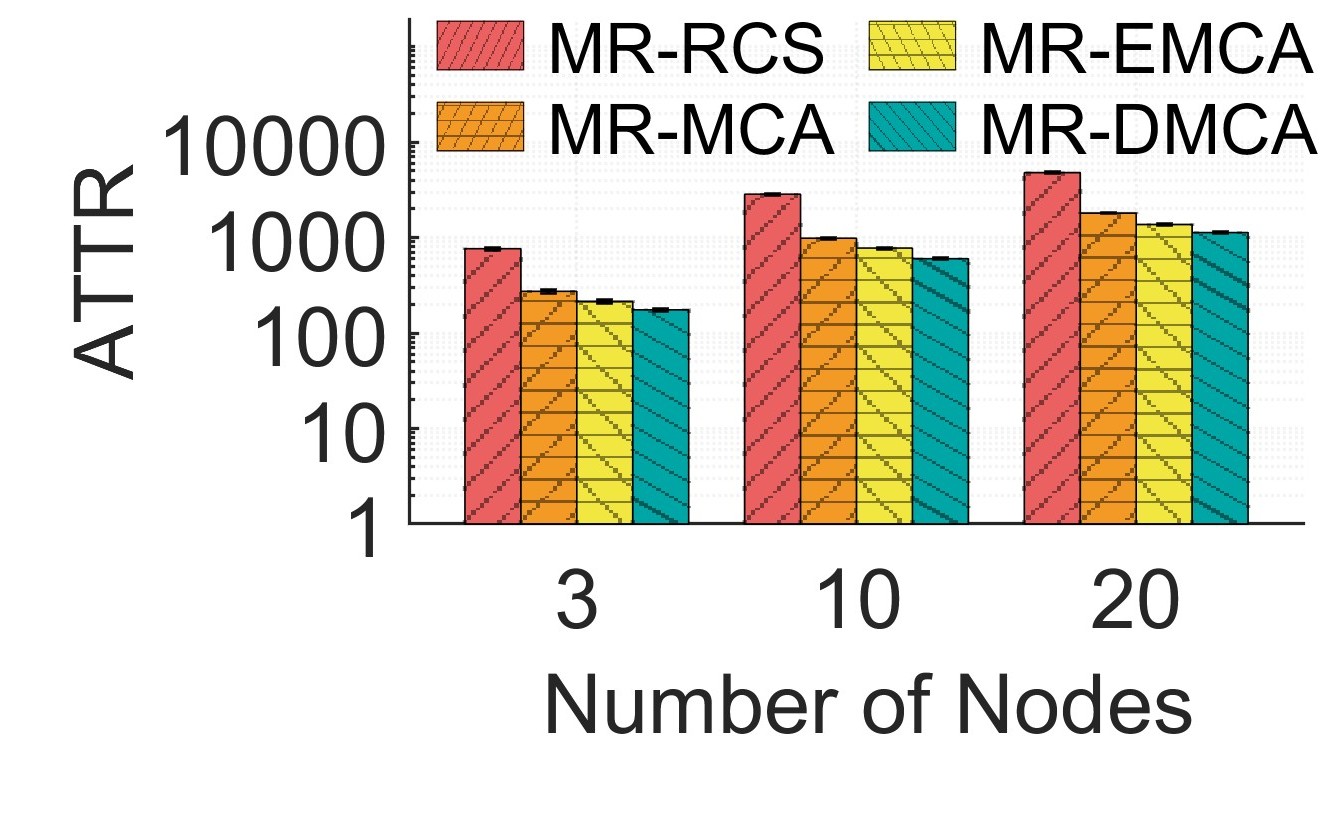}
			};
		\end{tikzpicture}\\[-0.25em]
		{\footnotesize (a)}
	\end{minipage}%
	\begin{minipage}[b]{0.5\columnwidth}
		\centering
		\begin{tikzpicture}
			\node[draw, line width=0.5pt, inner sep=2pt] {%
				\includegraphics[width=1.4in]{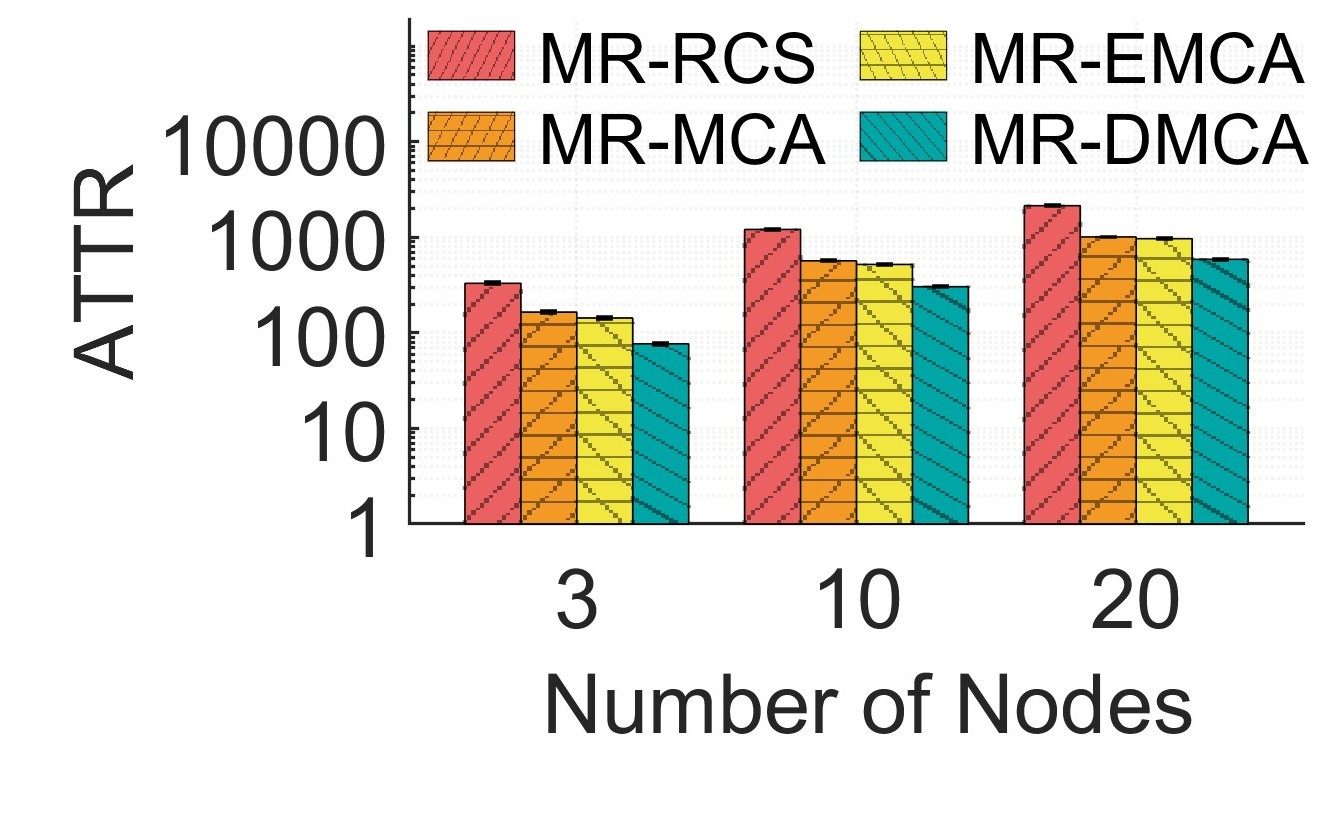}%
			};
		\end{tikzpicture}\\[-0.25em]
		{\footnotesize (b)}
	\end{minipage}
	\caption{Asym 20-CH (a) m=2 (b) m=5  with 85\%PR}\hfill
	\label{fig:IDN_R7cc_R8cc}
\end{figure}

To further evaluate the scalability of the proposed protocol, additional simulations were conducted under the controlled termination scenario, where all protocols employ both N-1 and IDN=0 conditions before terminating. In this experiment, the network size is increased to 20 nodes, and the asymmetric channel set is expanded to 20 channels (Asym 20-CH). The simulations consider channel similarity indices m=2 and m=5 under 0\% and 85\% PR activity.
The ATTR results for this scenario are presented in Fig.~\ref{fig:IDN_R7aa_R8aa} and Fig.~\ref{fig:IDN_R7cc_R8cc}. Since all protocols operate under the same termination condition, 100\% ATM is achieved for all cases, ensuring correct topology discovery across the network. Therefore, this experiment primarily evaluates the rendezvous efficiency and scalability of the protocols as the network size and channel search space increase.
To highlight the robustness of the proposed protocol, a worst-case configuration is considered where 20 nodes operate over 20 asymmetric channels with the lowest similarity index (m=2) under high PR activity (85\%). Under this challenging condition, MR-DMCA achieves an ATTR improvement of approximately 76\%, 37\%, and 17\% compared to MR-RCS, MR-MCA, and MR-EMCA, respectively.

These results demonstrate that MR-DMCA maintains efficient rendezvous performance even in large-scale cognitive radio networks with expanded channel sets and high primary user activity.

\section{Conclusion}
This paper presented MR-DMCA, a multihop reliable rendezvous protocol designed to address the challenges of incorrect neighbour discovery, premature termination, and incomplete topology formation in cognitive radio networks. The proposed approach incorporates coordinate-assisted validation and a controlled termination mechanism based on both the N-1 and IDN = 0 conditions to ensure correct neighbour discovery before terminating the rendezvous process.
Simulation results demonstrate that while conventional protocols may terminate earlier, they often suffer from incomplete topology discovery. In contrast, MR-DMCA guarantees 100\% topology correctness while maintaining efficient rendezvous performance. Under challenging conditions with low channel similarity (m=2) and high primary user activity, MR-DMCA achieves up to 76\%, 37\%, and 17\% improvement in ATTR compared to the baseline protocols while eliminating premature termination.
The results confirm that incorporating topology validation into the rendezvous process significantly improves the reliability of multihop neighbour discovery. As future work, the proposed approach will be extended to large-scale post-disaster cognitive radio networks, where nodes may operate under dynamic environments and limited knowledge of neighbouring nodes.
 
\FloatBarrier


\begin{thebibliography}{00}
	
	
	\bibliographystyle{plain}
	
	\bibitem{4} Federal Communications Commission,\url{http://www.it.kth.se/~jmitola/Mitola_Dissertation8_Integrated.pdf}, 2003. 
	
	\bibitem{1} 	S. Ghafoor, P. D. Sutton, C. J. Sreenan, and K. N. Brown, “Cognitive radio for disaster response networks: Survey, potential, and challenges,” IEEE Wirel. Commun., vol. 21, no. 5, pp. 70–80, 2014.
	
	\bibitem{2} 	M. Matracia, N. Saeed, M. A. Kishk, and M. S. Alouini, “Post-Disaster Communications: Enabling Technologies, Architectures, and Open Challenges,” IEEE Open J. Commun. Soc., vol. 3, 2022.
	
	\bibitem{3} 	G. Baldini, S. Karanasios, D. Allen, and F. Vergari, “Survey of wireless communication technologies for public safety,” IEEE Communications Surveys and Tutorials, vol. 16, no. 2. Institute of Electrical and Electronics Engineers Inc., 2014. 
	
	
	
	\bibitem{5} 	N. C. Theis, R. W. Thomas, and L. A. DaSilva, “Rendezvous for cognitive radios,” IEEE Trans. Mob. Comput., vol. 10, no. 2, pp. 216–227, Feb. 2011.
	
	
	\bibitem{Ali2025}
	Z. Ali, S. Ghafoor, S. Unnikrishnan, E. Furey, and I. McLoughlin, 
	“A dual modular clock algorithm for cognitive radio-based emergency response network,” 
	in \textit{Proceedings of the 2025 IEEE 22nd Consumer Communications \& Networking Conference (CCNC)}, 
	pp. 1--6, IEEE, 2025.
	
	\bibitem{Ali2026} Z. Ali, S. Unnikrishnan, E. Furey, I. McLoughlin, S. Ghafoor,
	“A Multihop Rendezvous Protocol for Cognitive Radio-based Emergency Response Network,”	arXiv preprint arXiv:2602.16367 [cs.NI].

		
	
	\bibitem{6} 	S. Ghafoor, C. J. Sreenan, and K. N. Brown, “A cognitive radio-based fully blind multihop rendezvous protocol for unknown environments,” Ad Hoc Networks, vol. 107, Oct. 2020.
	
	
	
	
	\bibitem{gg} I. F. Akyildiz, W.-Y. Lee, and K. R. Chowdhury, “Crahns: Cognitive radio ad hoc networks,” \textit{Ad Hoc Networks}, vol. 7, no. 5, 2009.
	
	\bibitem{hh} A. S. A. Ukey and M. Chawla, “Rendezvous in cognitive radio ad hoc networks: A survey,” \textit{International Journal of Ad Hoc and Ubiquitous Computing}, vol. 29, no. 4, 2018.
	
	
	
	\bibitem{li2017} A. Li, G. Han, J. J. P. C. Rodrigues, and S. Chan, "Channel Hopping Protocols for Dynamic Spectrum Management in 5G Technology," \emph{IEEE Wireless Communications}, vol. 24, 2017.
	
	
	\bibitem{Hliu2012}
	H. Liu et al.,
	``Jump-Stay Rendezvous Algorithm for Cognitive Radio Networks,''
	\emph{IEEE Transactions on Parallel and Distributed Systems}, vol. 23, no. 10, pp. 1867--1881, Oct. 2012.
	
	\bibitem{paul2016}
	R. Paul and Y. Choi,
	``Adaptive Rendezvous for Heterogeneous Channel Environments in Cognitive Radio Networks,''
	\emph{IEEE Transactions on Wireless Communications}, vol. 15, no. 11, pp. 7753--7765, Nov. 2016.
	
	\bibitem{yu2015}
	L. Yu et al.,
	``Multiple Radios for Fast Rendezvous in Cognitive Radio Networks,''
	\emph{IEEE Transactions on Mobile Computing}, vol. 14, no. 9, pp. 1917--1931, Sept. 2015.
	
	\bibitem{g1}
	W. Chen et al.,
	``Construction and Analysis of Shift-Invariant, Asynchronous-Symmetric Channel-Hopping Sequences for Cognitive Radio Networks,'' \emph{IEEE Transactions on Communications}, vol. 65, no. 4, pp. 1494--1506, Apr. 2017.
	
	\bibitem{g2}
	C. Chang, W. Liao, and T. Wu,
	``Tight Lower Bounds for Channel Hopping Schemes in Cognitive Radio Networks,'' \emph{IEEE/ACM Transactions on Networking}, vol. 24, no. 4, pp. 2343--2356, Aug. 2016.
	
	
	
	\bibitem{chuang2015}
	I. Chuang, H. Wu, and Y. Kuo,
	``A Fast Rendezvous-Guarantee Channel Hopping Protocol for Cognitive Radio Networks,''
	\emph{IEEE Transactions on Vehicular Technology}, vol. 64, no. 12, pp. 5804--5816, Dec. 2015.
	
	\bibitem{chao2016}
	C. Chao and H. Fu,
	``Supporting Fast Rendezvous Guarantee by Randomized Quorum and Latin Square for Cognitive Radio Networks,''
	\emph{IEEE Transactions on Vehicular Technology}, vol. 65, no. 10, pp. 8388--8399, Oct. 2016.
	
	\bibitem{sheu2016}
	J. Sheu, C. Su, and G. Chang,
	``Asynchronous Quorum-Based Blind Rendezvous Schemes for Cognitive Radio Networks,''
	\emph{IEEE Transactions on Communications}, vol. 64, no. 3, pp. 918--930, Mar. 2016.
	
	\bibitem{zhang2014}
	Y. Zhang et al.,
	``Channel-Hopping-Based Communication Rendezvous in Cognitive Radio Networks,''
	\emph{IEEE/ACM Transactions on Networking}, vol. 22, no. 3, pp. 889--902, Jun. 2014.
	
	
	\bibitem{huang2017}
	J. Huang, G. Chang, and J. Huang,
	``Anti-Jamming Rendezvous Scheme for Cognitive Radio Networks,''
	\emph{IEEE Transactions on Mobile Computing}, vol. 16, no. 3, pp. 648--661, Mar. 2017.
	
	\bibitem{chang2015}
	G. Chang, J. Huang, and Y. Wang,
	``Matrix-Based Channel Hopping Algorithm for Cognitive Radio Networks,''
	\emph{IEEE Transactions on Wireless Communications}, vol. 14, no. 5, pp. 2755--2768, May 2015.
	
	
	\bibitem{sahoo2016}
	P. Sahoo and D. Sahoo,
	``Sequence-Based Channel Hopping Algorithms for Dynamic Spectrum Sharing in Cognitive Radio Networks,''
	\emph{IEEE Journal on Selected Areas in Communications}, vol. 34, no. 11, pp. 2814--2828, Nov. 2016.
	
	\bibitem{chao2016efficient}
	C. Chao, C. Hsu, and Y. Ling,
	``Efficient Asynchronous Channel Hopping Design for Cognitive Radio Networks,''
	\emph{IEEE Transactions on Vehicular Technology}, vol. 65, no. 9, pp. 6888--6900, Sept. 2016.
	
	\bibitem{chang2014}
	G. Chang et al.,
	``Novel Channel-Hopping Schemes for Cognitive Radio Networks,''
	\emph{IEEE Transactions on Mobile Computing}, vol. 13, no. 2, pp. 407--421, Feb. 2014.
	
	\bibitem{yang2016}
	B. Yang et al.,
	``Fully Distributed Channel-Hopping Algorithms for Rendezvous Setup in Cognitive Multi-Radio Networks,''
	\emph{IEEE Transactions on Vehicular Technology}, vol. 65, no. 10, pp. 8629--8643, Oct. 2016.
	
	
	\bibitem{ii} 	J. Shin, D. Yang, and C. Kim, “A channel rendezvous scheme for cognitive radio networks,” IEEE Commun. Lett., vol. 14, 2010.
	
	
	\bibitem{jj} 	C. M. Chao, L. F. Lien, and C. Y. Hsu, “Rendezvous enhancement in arbitrary-duty-cycled wireless sensor networks,” IEEE Trans. Wirel. Commun., vol. 12, no. 8, pp. 4080–4091, 2013, doi: 10.1109/TCOMM.2013.051313.121688.
	
	\bibitem{kk} 	J. P. Sheu and J. J. Lin, “A Multi-Radio Rendezvous Algorithm Based on Chinese Remainder Theorem in Heterogeneous Cognitive Radio Networks,” IEEE Trans. Mob. Comput., vol. 17, 2018.
	
	\bibitem{ll} 	Y. C. Chang, C. S. Chang, and J. P. Sheu, “An Enhanced Fast Multi-Radio Rendezvous Algorithm in Heterogeneous Cognitive Radio Networks,” IEEE Trans. Cogn. Commun. Netw., vol. 4, 2018.
	
	\bibitem{mm} 	M. T. Islam, S. Kandeepan, and R. J. Evans, “Prime Number Theory based Multi-Radio Rendezvous for Cognitive Radio Communication,” 2019 2nd IEEE Int. Conf. Inf. Commun. Signal Process. ICICSP 2019.
	
	\bibitem{oo} 	Z. Zhixin, Y. Deng, Z. Xiaohong, Z. Xianfei, H. Liqin, and Z. Zhidong, “Multiple Prime Expansion Channel Hopping for Blind Rendezvous in a Wireless Sensor Network,” Wirel. Commun. Mob. Comput., 2022.
	
	\bibitem{pp} 	Z. Gu, Y. Wang, T. Shen, and F. C. M. Lau, “On heterogeneous sensing capability for distributed rendezvous in cognitive radio networks,” IEEE Trans. Mob. Comput., vol. 20, 2021.
	
	
	
	
	
	
	
	\bibitem{13} H. Liu, Z. Lin, X. Chu, and Y. W. Leung, “Jump-stay rendezvous algorithm for cognitive radio networks,” *IEEE Transactions on Parallel and Distributed Systems*, vol. 23, no. 10, pp. 1867--1881, IEEE, 2012.
	
	
	\bibitem{15}
	D. D. Onthoni, P. K. Sahoo, and M. Atiquzzaman, ``ASAA: Multihop and Multiuser Channel Hopping Protocols for Cognitive-Radio-Enabled Internet of Things,'' \textit{IEEE Internet of Things Journal}, vol.~10, no.~9, pp.~8305--8318, 2023.
	
	\bibitem{16}
	J. Jia and Q. Zhang, ``Rendezvous protocols based on message passing in cognitive radio networks,'' \textit{IEEE Transactions on Wireless Communications}, vol.~12, no.~11, pp.~5594--5606, 2013.

	
	
	\bibitem{rr} Z. Gu, T. Shen, Y. Wang, and F. C. Lau, “Efficient rendezvous for heterogeneous interference in cognitive radio networks,” IEEE Trans. Wireless Commun., vol. 19, no. 1, 2019.
	
	\bibitem{ss} M. Yuan, Y. Chu, and W. Guo, “Frequency-Gateway Based Differential Rendezvous Algorithm for Cognitive Radio Networks,” in \textit{Proc. IEEE Wireless Commun. Netw. Conf. (WCNC)}, Apr. 2024.
	
	\bibitem{tt} P. D. Sutton, K. E. Nolan, and L. E. Doyle, “Cyclostationary signatures in practical cognitive radio applications,” J. Sel. Areas Commun., vol. 26, 2008.
	
	\bibitem{bk} H. Kim and K. Shin, “Efficient discovery of spectrum opportunities with MAC-layer sensing in cognitive radio networks,” IEEE Trans. Mob. Comput., vol. 7, 2008.
	
	\bibitem{uu} Y. Saleem and M. H. Rehmani, “Primary radio user activity models for cognitive radio networks: A survey,” Journal of Network and Computer Applications, vol. 43,  2014.
	

	
	\bibitem{bj} H. Kim and K. Shin, “Fast discovery of spectrum opportunities in cognitive radio networks,” in IEEE DySPAN, 2008.
	
	\bibitem{xx} A. W. Min and K. G. Shin, “Exploiting multi-channel diversity in spesctrum-agile networks,” in Proceedings of the INFOCOM, 2008.
	
	
	
	\bibitem{Clark1990}
	B.~N. Clark, C.~J. Colbourn, and D.~S. Johnson, ``Unit disk graphs,'' \emph{Discrete Mathematics}, 86(1–3):165--177, 1990.
	\bibitem{Gilbert1961}
	E.~N. Gilbert, ``Random plane networks,'' \emph{SIAM J.}, 9(4):533--543, 1961.
	\bibitem{Penrose2003}
	M.~Penrose, \emph{Random Geometric Graphs}. Oxford Univ. Press, 2003.
	\bibitem{Kuhn2008}
	F.~Kuhn, R.~Wattenhofer, and A.~Zollinger, ``Ad hoc networks beyond unit disk graphs,'' \emph{Wireless Networks}, 14(5):715--729, 2008.
	

	\bibitem{chigan} 
	Tricia Chigan, \emph{CRCN Simulator}, 
	Available at: \url{https://faculty.uml.edu/Tricia_Chigan/Research/CRCN_Simulator.htm}, 
	2024.
	
	
	
	\bibitem{bonnmotion_web}
	BonnMotion,
	``BonnMotion: A Mobility Scenario Generation and Analysis Tool,''
	[Online]. Available: \url{https://bonnmotion.sys.cs.uos.de/index.shtml}.
	
	
	
	
	
\end{thebibliography}
\end{document}